\documentclass[a4papers,runningheads]{llncs}

\usepackage{hyperref}

\usepackage{cite}

\newif\ifblind
\blindfalse

\newif\ifdraft
\drafttrue

\newif\iflong
\longfalse

\usepackage[T1]{fontenc}
\usepackage{times}
\usepackage[scaled=0.81]{beramono}

\iflong
\newcommand{\nicepar}[1]{\paragraph{#1}}
\else
\makeatletter
\newcommand\nicepar{\@startsection{paragraph}{4}{\z@}%
  {2\p@ \@plus -4\p@ \@minus -4\p@}%
  {-0.5em \@plus -0.22em \@minus -0.1em}%
  {\bfseries\normalsize}}
\newcommand\nicesubsubsection{\@startsection{subsubsection}{3}{\z@}%
  {6\p@ \@plus -0.5\p@ \@minus -0.5\p@}%
  {-0.5em \@plus -0.22em \@minus -0.1em}%
  {\normalfont\normalsize\bfseries\boldmath}}
\makeatother
\renewcommand{\subsubsection}[1]{\nicesubsubsection{#1.}}
\fi

\usepackage{graphicx}

\usepackage{tikz}
\usetikzlibrary{shapes,arrows,positioning,calc,fit,intersections,through}

\definecolor{acol}{HTML}{88a0dc}
\definecolor{bcol}{HTML}{381a61}
\definecolor{ccol}{HTML}{7c4b73}
\definecolor{dcol}{HTML}{ed968c}
\definecolor{ecol}{HTML}{ab3329}
\definecolor{fcol}{HTML}{e78429}
\definecolor{gcol}{HTML}{f9d14a}
\definecolor{hcol}{HTML}{04d17c}

\colorlet{bbcol}{fcol}
\colorlet{boogiecol}{dcol}
\colorlet{jimplecol}{acol}
\colorlet{bytecodecol}{gcol}
\colorlet{jmlcol}{hcol}

\usepackage[inline]{enumitem}
\setlist[enumerate]{label=\emph{\roman*})}

\usepackage{ragged2e}

\usepackage{booktabs}
\usepackage{relsize}
\usepackage{changepage}            
\usepackage[skip=1pt]{caption}
\usepackage[skip=1pt]{subcaption}  

\graphicspath{{./figs/}}
\DeclareGraphicsExtensions{.pdf,.png}

\usepackage{amsmath,amsfonts,amssymb}
\usepackage{mathtools,xstring}
\usepackage{array}

\usepackage{listings}

\usepackage{pgfkeys,pgfplots,numprint,relsize,refcount}
\def\NAN{??}              
\def\keyfamily{/bbe/}

\usepackage{fmtcount}     

\DeclareDocumentCommand{\spellout}{m O{13}}{%
  \IfStrEq{#1}{ }{\NAN}{
    \IfInteger{#1}{
      \expandafter\ifnum#1<#2{\numberstringnum{#1}}\else#1\fi%
    }{\NAN}
}}

\DeclareDocumentCommand{\n}{t. t: o m o O{} t| t!}{%
  \begingroup
  \pgfkeys{/pgf/fpu=true}
  \IfBooleanTF{#2}{%
    \pgfkeyssetvalue{/tmp/value}{#4}%
    \pgfkeyssetvalue{/tmp/found}{found}%
  }{
    \pgfkeysifdefined{\keyfamily#4}{%
      \pgfkeyssetvalue{/tmp/value}{\pgfkeysvalueof{\keyfamily#4}}%
      \pgfkeyssetvalue{/tmp/found}{found}%
    }{}
  }%
  \pgfkeysifdefined{/tmp/found}{%
    \IfNoValueF{#5}{%
      \pgfkeyssetvalue{/tmp/multiplier}{#5}%
      \pgfmathparse{\pgfkeysvalueof{/tmp/multiplier} * \pgfkeysvalueof{/tmp/value}}%
      \pgfkeyslet{/tmp/value}\pgfmathresult%
    }%
    \IfBooleanTF{#1}{%
      \IfBooleanTF{#8}{%
        \spellout{\pgfkeysvalueof{/tmp/value}}}{%
        \pgfkeysvalueof{/tmp/value}}%
    }{
      \IfNoValueTF{#3}{%
        \pgfmathprintnumber%
        [set thousands separator={\,},int detect,#6]%
        {\pgfkeysvalueof{/tmp/value}}%
      }{
        {\pgfmathprintnumber%
          [precision=#3,fixed,zerofill,set thousands separator={\,},#6]%
          {\pgfkeysvalueof{/tmp/value}}}%
      }%
    }%
    \IfBooleanT{#7}{{\smaller[1.2]\%}}%
  }{
    \NAN%
  }%
  \pgfkeys{/pgf/fpu=false}
  \endgroup%
}

\lstdefinestyle{displayed}{
  numbers=left, %
  firstnumber=last, 
  frame=single, %
  breaklines=true,
  numbersep=2pt,
  frame=tb,
  numberstyle=\tiny,
  tabsize=2,
  captionpos=b,
  xleftmargin=0mm, %
  xrightmargin=0mm, %
  basicstyle=\ttfamily\scriptsize,
  keywordstyle=\bfseries\ttfamily,
  commentstyle=\color{darkgray}\itshape\ttfamily,
  keepspaces=true,
  columns=fixed,
  escapeinside={(*}{*)},
  mathescape=true,
  showstringspaces=false
}

\lstdefinestyle{plain}{
  numbers=none, %
  frame=none, %
  breaklines=false,
  tabsize=2,
  xleftmargin=2mm, %
  xrightmargin=2mm, %
  basicstyle=\ttfamily\scriptsize,
  keywordstyle=\bfseries\ttfamily,
  commentstyle=\color{darkgray}\itshape\ttfamily,
  keepspaces=true,
  columns=fullflexible,
  escapeinside={(*}{*)},
  mathescape=true,
  showstringspaces=false
}

\lstdefinelanguage{JavaRecent}[]{Java}
{
  morekeywords={var,yield}
}

\usepackage{boogie}

\lstdefinelanguage{Jimple}[]{JVMIS}
{
  morekeywords=[2]{if,throw,@caught},
  morekeywords=[4]{head,back,exit,hE,skip,normal},
  literate=%
  {:=}{$:$=}2
  ,
   mathescape=true,
   escapeinside={(*}{*)},
   identifierstyle=\ttfamily,
   keywordstyle=[2]{\bfseries\ttfamily},
   keywordstyle=[3]{\color{jimplecol}\bfseries\ttfamily},
   keywordstyle=[4]{\itshape\underbar},
}

\lstdefinelanguage{Vimp}[]{Jimple}
{
  morekeywords=[2]{assume,assert,invariant,@thrown},
  deletekeywords=[2]{@caught},
  literate=%
    {==}{$=$}1
    {!}{$\lnot\ $}1
    {!=}{$\neq$}1
    {||}{$\lor\;$}2
    {&&}{$\land\;$}2
    {==>}{$\,\Longrightarrow\;$}4
    {forall}{$\forall$}1
    {exists}{$\exists$}1
    {:}{$\colon$}1
    {::}{$\colon\!\!\colon$}1
    {:=}{$:$=}2
    {<}{$<$}1
    {<=}{$\le$}1
    {>}{$>$}1
    {>=}{$\ge$}1
}

\lstdefinelanguage[JML]{Java}[]{JavaRecent}%
       {
        comment=[l]{//\ },
        morecomment=[s]{/*\ }{*/},        
        morecomment=[s]{/**}{*/},
        classoffset=1,
        morekeywords={abrupt_behavior,abrupt_behaviour,
         accessible,accessible_redundantly,also,assert,assert_redundantly,
         assignable,assignable_redundantly,assume,assume_redundantly,
         axiom,behavior,behaviour,breaks,breaks_redundantly,
         callable,callable_redundantly,captures,captures_redundantly,
         choose,choose_if,code,code_bigint_math,code_java_math,
         code_safe_math,constraint,constraint_redundantly,constructor,
         continues,continues_redundantly,decreases,decreases_redundantly,
         decreasing,decreasing_redundantly,diverges,diverges_redundantly,
         duration,duration_redundantly,ensures,ensures_redundantly,
         example,exceptional_behavior,exceptional_behaviour,
         exceptional_example,exsures,exsures_redundantly,extract,field,
         forall,for_example,ghost,helper,hence_by,hence_by_redundantly,
         implies_that,in,in_redundantly,initializer,initially,instance,
         invariant,invariant_redundantly,loop_invariant,
         loop_invariant_redundantly,maintaining,maintaining_redundantly,
         maps,maps_redundantly,measured_by,measured_by_redundantly,method,
         model,model_program,modifiable,modifiable_redundantly,modifies,
         modifies_redundantly,monitored,monitors_for,non_null,
         normal_behavior,normal_behaviour,normal_example,nowarn,
         nullable,nullable_by_default,old,or,post,post_redundantly,
         pre,pre_redundantly,pure,readable,refine,refines,refining,represents,
         represents_redundantly,requires,requires_redundantly,
         returns,returns_redundantly,set,signals,signals_only,
         signals_only_redundantly,signals_redundantly,spec_bigint_math,
         spec_java_math,spec_protected,spec_public,spec_safe_math,
         static_initializer,uninitialized,unreachable,weakly,
         when,when_redundantly,working_space,working_space_redundantly,
         writable
        },
        morekeywords={rep,peer,readonly},
        keywordsprefix=\\,
        otherkeywords={<:,<-,->,..,<==,==>,<==>,<=!=>},
        keywordstyle={\color{jmlcol}\bfseries\ttfamily},
        classoffset=0, 
        keywordstyle={\bfseries\ttfamily}
}

\lstdefinelanguage{BBlib}[]{JavaRecent}
{
  morekeywords={forall,exists,old},
  morekeywords=[2]{invariant,assertion,assumption},
  morekeywords=[3]{@Require,@Ensure,@Raise,@Return,@Predicate,@Attach},
  deletekeywords=[2]{int,float},
  literate=%
  {:}{$\colon$}1
  {::}{$\bullet$}1
  {!}{$\lnot\ $}1
  {==}{$=$}1
  {!=}{$\neq$}1
  {&&}{$\land$}1
  {||}{$\lor$}1
  {<}{$<$}1
  {<=}{$\le$}1
  {>}{$>$}1
  {>=}{$\ge$}1
  {==>}{$\Longrightarrow$}3
  {<==>}{$\Longleftrightarrow$}4
  {\\forall}{$\forall$}1
  {\\exists}{$\exists$}1
  ,
  mathescape=true,
  escapeinside={(*}{*)},
  identifierstyle=\ttfamily,
  keywordstyle=[2]{\color{bbcol}\bfseries\ttfamily},
  keywordstyle=[3]{\color{bbcol}\bfseries\ttfamily},
}
 
\newcommand{\J}[1]{\mbox{\lstinline[basicstyle=\ttfamily,language=JavaRecent]|#1|}}

\newcommand{\BBl}[1]{\mbox{\lstinline[basicstyle=\ttfamily,language=BBlib]|#1|}}
\newcommand{\JML}[1]{\mbox{\lstinline[basicstyle=\ttfamily,language={[JML]Java}]|#1|}}

\newcommand{\Gr}[1]{\mbox{\lstinline[basicstyle=\ttfamily,language=Jimple]|#1|}}
\newcommand{\Vi}[1]{\mbox{\lstinline[basicstyle=\ttfamily,language=Vimp]|#1|}}

\usepackage{multicol,multirow}
\usepackage{xurl}
\usepackage{xspace}
\usepackage{fontawesome}

\usepackage{savesym}
\savesymbol{endnote}

\usepackage{enotez}

\restoresymbol{enotez}{endnote}

\setenotez{list-name={URL References}}
\setenotez{counter-format=arabic}
\setenotez{mark-cs={\formatEndNoteMark}}
\newcommand{\formatEndNoteMark}[1]{\textsuperscript{\{\color{blue}{\textsf{#1}}\}}}

\DeclareInstance{enotez-list}{custom}{list}%
{list-type=itemize,heading=\section*{#1}}

\DeclareDocumentCommand{\urlcite}{s O{} m}
{\enotezendnote{\IfNoValueF{#2}{#2\xspace}\url{#3}}}

\newcommand{\byteback}{{\smaller[0.5]{\textsc{Byte\-Back}}}\xspace}
\newcommand{\bblib}{\texttt{BBlib}\xspace}

\DeclareDocumentCommand{\op}{o}{%
  \ensuremath{\IfNoValueTF{#1}{\Vi{op}}{#1_{\Vi{op}}}}%
  \xspace
}

\DeclareDocumentCommand{\tr}{s o}{%
  \ensuremath{\mathcal{V}\IfNoValueF{#2}{_{\mathsf{#2}}}}\xspace
}

\DeclareMathOperator{\subtype}{\preceq}
\DeclareMathOperator{\notsubtype}{\npreceq}
\DeclareMathOperator{\sand}{\;\land\;}

\newcounter{experimentrowcounter}
\newcommand\exprowc{\refstepcounter{experimentrowcounter}\arabic{experimentrowcounter}}

\newcommand\exprow[1]{
    \n[1]{#1/ConversionTime}[0.001]
    & \n[1]{#1/VerificationTime}[0.001]
    & \n[0]{#1/SourceLinesOfCode}
    & \n[0]{#1/BoogieLinesOfCode} 
    & \n[0]{#1/MethodCount}
    & \n[0]{#1/SpecPredicateCount}
    & \n[0]{#1/SpecFunctionalCount}
    & \n[0]{#1/SpecExceptionCount} \\
}

\usepackage{ifthen}

\ifdraft
\usepackage[colorinlistoftodos,textsize=scriptsize,obeyFinal,textwidth=33mm]{todonotes}

\DeclareDocumentCommand{\ReviewNote}{s o m O{white}}{%
  \todo[color=#4,\IfBooleanTF{#1}{inline}{}]{\IfNoValueF{#2}{\textbf{#2:}\xspace}#3}
}
\else
\DeclareDocumentCommand{\ReviewNote}{s o m O{white}}{}
\fi

\DeclareDocumentCommand{\caf}{s m}{\IfBooleanTF{#1}{\ReviewNote*{#2}[yellow]}{\ReviewNote{#2}[yellow]}}
\DeclareDocumentCommand{\mpg}{s m}{\IfBooleanTF{#1}{\ReviewNote*{#2}[red!70!white]}{\ReviewNote{#2}[red!70!white]}}

\pgfkeyssetvalue{/bb/number of experiments}{34}

\begin{document}

\title{Reasoning About Exceptional Behavior \\At the Level of Java Bytecode\ifblind\else\thanks{Work partially supported by SNF grant 200021-207919 (LastMile).}\fi}

\titlerunning{Reasoning About Bytecode Exceptional Behavior}

\pgfkeyssetvalue{/bbe/count/non-exceptional}{35}
\pgfkeyssetvalue{/bbe/count/java}{20}
\pgfkeyssetvalue{/bbe/count}{37}
\pgfkeyssetvalue{/bbe/count/j8}{17}
\pgfkeyssetvalue{/bbe/count/j17}{3}
\pgfkeyssetvalue{/bbe/count/s2}{9}
\pgfkeyssetvalue{/bbe/count/k18}{8}
\pgfkeyssetvalue{/bbe/count/method}{1070}
\pgfkeyssetvalue{/bbe/count/method/j8}{884}
\pgfkeyssetvalue{/bbe/count/method/j17}{25}
\pgfkeyssetvalue{/bbe/count/method/s2}{84}
\pgfkeyssetvalue{/bbe/count/method/k18}{77}
\pgfkeyssetvalue{/bbe/count/raises}{80}
\pgfkeyssetvalue{/bbe/count/raises/j8}{56}
\pgfkeyssetvalue{/bbe/count/raises/j17}{4}
\pgfkeyssetvalue{/bbe/count/raises/s2}{10}
\pgfkeyssetvalue{/bbe/count/raises/k18}{10}
\pgfkeyssetvalue{/bbe/count/returns}{119}
\pgfkeyssetvalue{/bbe/count/returns/j8}{66}
\pgfkeyssetvalue{/bbe/count/returns/j17}{5}
\pgfkeyssetvalue{/bbe/count/returns/s2}{25}
\pgfkeyssetvalue{/bbe/count/returns/k18}{23}
\pgfkeyssetvalue{/bbe/count/invariants}{45}
\pgfkeyssetvalue{/bbe/count/invariants/j8}{27}
\pgfkeyssetvalue{/bbe/count/invariants/j17}{2}
\pgfkeyssetvalue{/bbe/count/invariants/s2}{8}
\pgfkeyssetvalue{/bbe/count/invariants/k18}{8}
\pgfkeyssetvalue{/bbe/count/assertions}{86}
\pgfkeyssetvalue{/bbe/count/assertions/j8}{59}
\pgfkeyssetvalue{/bbe/count/assertions/j17}{2}
\pgfkeyssetvalue{/bbe/count/assertions/s2}{13}
\pgfkeyssetvalue{/bbe/count/assertions/k18}{12}
\pgfkeyssetvalue{/bbe/j8/exceptions/PotentialIndexOutOfBounds/ConversionTime}{1189.2}
\pgfkeyssetvalue{/bbe/j8/exceptions/PotentialIndexOutOfBounds/ConversionOverhead}{0.094149590859368}
\pgfkeyssetvalue{/bbe/j8/exceptions/PotentialIndexOutOfBounds/VerificationTime}{977.0}
\pgfkeyssetvalue{/bbe/j8/exceptions/PotentialIndexOutOfBounds/SourceLinesOfCode}{87}
\pgfkeyssetvalue{/bbe/j8/exceptions/PotentialIndexOutOfBounds/BytecodeLinesOfCode}{147}
\pgfkeyssetvalue{/bbe/j8/exceptions/PotentialIndexOutOfBounds/BoogieLinesOfCode}{366}
\pgfkeyssetvalue{/bbe/j8/exceptions/PotentialIndexOutOfBounds/MethodCount}{16}
\pgfkeyssetvalue{/bbe/j8/exceptions/PotentialIndexOutOfBounds/SpecRequireCount}{2}
\pgfkeyssetvalue{/bbe/j8/exceptions/PotentialIndexOutOfBounds/SpecEnsureCount}{0}
\pgfkeyssetvalue{/bbe/j8/exceptions/PotentialIndexOutOfBounds/SpecRaiseCount}{5}
\pgfkeyssetvalue{/bbe/j8/exceptions/PotentialIndexOutOfBounds/SpecReturnCount}{3}
\pgfkeyssetvalue{/bbe/j8/exceptions/PotentialIndexOutOfBounds/SpecPredicateCount}{5}
\pgfkeyssetvalue{/bbe/j8/exceptions/PotentialIndexOutOfBounds/SpecPureCount}{2}
\pgfkeyssetvalue{/bbe/j8/exceptions/PotentialIndexOutOfBounds/SpecAssertionCount}{0}
\pgfkeyssetvalue{/bbe/j8/exceptions/PotentialIndexOutOfBounds/SpecAssumptionCount}{0}
\pgfkeyssetvalue{/bbe/j8/exceptions/PotentialIndexOutOfBounds/SpecInvariantCount}{0}
\pgfkeyssetvalue{/bbe/j8/exceptions/PotentialIndexOutOfBounds/SpecExceptionCount}{8}
\pgfkeyssetvalue{/bbe/j8/exceptions/PotentialIndexOutOfBounds/SpecFunctionalCount}{2}
\pgfkeyssetvalue{/bbe/j8/exceptions/PotentialIndexOutOfBounds/SpecIntermediateCount}{0}
\pgfkeyssetvalue{/bbe/j8/exceptions/PotentialNullDereference/ConversionTime}{1009.8}
\pgfkeyssetvalue{/bbe/j8/exceptions/PotentialNullDereference/ConversionOverhead}{0.0971795367816603}
\pgfkeyssetvalue{/bbe/j8/exceptions/PotentialNullDereference/VerificationTime}{959.6}
\pgfkeyssetvalue{/bbe/j8/exceptions/PotentialNullDereference/SourceLinesOfCode}{84}
\pgfkeyssetvalue{/bbe/j8/exceptions/PotentialNullDereference/BytecodeLinesOfCode}{230}
\pgfkeyssetvalue{/bbe/j8/exceptions/PotentialNullDereference/BoogieLinesOfCode}{429}
\pgfkeyssetvalue{/bbe/j8/exceptions/PotentialNullDereference/MethodCount}{26}
\pgfkeyssetvalue{/bbe/j8/exceptions/PotentialNullDereference/SpecRequireCount}{0}
\pgfkeyssetvalue{/bbe/j8/exceptions/PotentialNullDereference/SpecEnsureCount}{0}
\pgfkeyssetvalue{/bbe/j8/exceptions/PotentialNullDereference/SpecRaiseCount}{2}
\pgfkeyssetvalue{/bbe/j8/exceptions/PotentialNullDereference/SpecReturnCount}{8}
\pgfkeyssetvalue{/bbe/j8/exceptions/PotentialNullDereference/SpecPredicateCount}{4}
\pgfkeyssetvalue{/bbe/j8/exceptions/PotentialNullDereference/SpecPureCount}{0}
\pgfkeyssetvalue{/bbe/j8/exceptions/PotentialNullDereference/SpecAssertionCount}{0}
\pgfkeyssetvalue{/bbe/j8/exceptions/PotentialNullDereference/SpecAssumptionCount}{0}
\pgfkeyssetvalue{/bbe/j8/exceptions/PotentialNullDereference/SpecInvariantCount}{0}
\pgfkeyssetvalue{/bbe/j8/exceptions/PotentialNullDereference/SpecExceptionCount}{10}
\pgfkeyssetvalue{/bbe/j8/exceptions/PotentialNullDereference/SpecFunctionalCount}{0}
\pgfkeyssetvalue{/bbe/j8/exceptions/PotentialNullDereference/SpecIntermediateCount}{0}
\pgfkeyssetvalue{/bbe/j8/exceptions/MultiCatch/ConversionTime}{1208.0}
\pgfkeyssetvalue{/bbe/j8/exceptions/MultiCatch/ConversionOverhead}{0.0781488241192683}
\pgfkeyssetvalue{/bbe/j8/exceptions/MultiCatch/VerificationTime}{928.6}
\pgfkeyssetvalue{/bbe/j8/exceptions/MultiCatch/SourceLinesOfCode}{67}
\pgfkeyssetvalue{/bbe/j8/exceptions/MultiCatch/BytecodeLinesOfCode}{102}
\pgfkeyssetvalue{/bbe/j8/exceptions/MultiCatch/BoogieLinesOfCode}{311}
\pgfkeyssetvalue{/bbe/j8/exceptions/MultiCatch/MethodCount}{10}
\pgfkeyssetvalue{/bbe/j8/exceptions/MultiCatch/SpecRequireCount}{0}
\pgfkeyssetvalue{/bbe/j8/exceptions/MultiCatch/SpecEnsureCount}{1}
\pgfkeyssetvalue{/bbe/j8/exceptions/MultiCatch/SpecRaiseCount}{0}
\pgfkeyssetvalue{/bbe/j8/exceptions/MultiCatch/SpecReturnCount}{4}
\pgfkeyssetvalue{/bbe/j8/exceptions/MultiCatch/SpecPredicateCount}{1}
\pgfkeyssetvalue{/bbe/j8/exceptions/MultiCatch/SpecPureCount}{0}
\pgfkeyssetvalue{/bbe/j8/exceptions/MultiCatch/SpecAssertionCount}{4}
\pgfkeyssetvalue{/bbe/j8/exceptions/MultiCatch/SpecAssumptionCount}{0}
\pgfkeyssetvalue{/bbe/j8/exceptions/MultiCatch/SpecInvariantCount}{0}
\pgfkeyssetvalue{/bbe/j8/exceptions/MultiCatch/SpecExceptionCount}{4}
\pgfkeyssetvalue{/bbe/j8/exceptions/MultiCatch/SpecFunctionalCount}{1}
\pgfkeyssetvalue{/bbe/j8/exceptions/MultiCatch/SpecIntermediateCount}{4}
\pgfkeyssetvalue{/bbe/j8/exceptions/Basic/ConversionTime}{1084.8}
\pgfkeyssetvalue{/bbe/j8/exceptions/Basic/ConversionOverhead}{0.0925370963990726}
\pgfkeyssetvalue{/bbe/j8/exceptions/Basic/VerificationTime}{1083.2}
\pgfkeyssetvalue{/bbe/j8/exceptions/Basic/SourceLinesOfCode}{164}
\pgfkeyssetvalue{/bbe/j8/exceptions/Basic/BytecodeLinesOfCode}{474}
\pgfkeyssetvalue{/bbe/j8/exceptions/Basic/BoogieLinesOfCode}{504}
\pgfkeyssetvalue{/bbe/j8/exceptions/Basic/MethodCount}{46}
\pgfkeyssetvalue{/bbe/j8/exceptions/Basic/SpecRequireCount}{0}
\pgfkeyssetvalue{/bbe/j8/exceptions/Basic/SpecEnsureCount}{0}
\pgfkeyssetvalue{/bbe/j8/exceptions/Basic/SpecRaiseCount}{10}
\pgfkeyssetvalue{/bbe/j8/exceptions/Basic/SpecReturnCount}{7}
\pgfkeyssetvalue{/bbe/j8/exceptions/Basic/SpecPredicateCount}{10}
\pgfkeyssetvalue{/bbe/j8/exceptions/Basic/SpecPureCount}{8}
\pgfkeyssetvalue{/bbe/j8/exceptions/Basic/SpecAssertionCount}{3}
\pgfkeyssetvalue{/bbe/j8/exceptions/Basic/SpecAssumptionCount}{0}
\pgfkeyssetvalue{/bbe/j8/exceptions/Basic/SpecInvariantCount}{0}
\pgfkeyssetvalue{/bbe/j8/exceptions/Basic/SpecExceptionCount}{17}
\pgfkeyssetvalue{/bbe/j8/exceptions/Basic/SpecFunctionalCount}{0}
\pgfkeyssetvalue{/bbe/j8/exceptions/Basic/SpecIntermediateCount}{3}
\pgfkeyssetvalue{/bbe/j8/exceptions/Loop/ConversionTime}{1124.8}
\pgfkeyssetvalue{/bbe/j8/exceptions/Loop/ConversionOverhead}{0.0929090416385523}
\pgfkeyssetvalue{/bbe/j8/exceptions/Loop/VerificationTime}{970.4}
\pgfkeyssetvalue{/bbe/j8/exceptions/Loop/SourceLinesOfCode}{97}
\pgfkeyssetvalue{/bbe/j8/exceptions/Loop/BytecodeLinesOfCode}{161}
\pgfkeyssetvalue{/bbe/j8/exceptions/Loop/BoogieLinesOfCode}{398}
\pgfkeyssetvalue{/bbe/j8/exceptions/Loop/MethodCount}{11}
\pgfkeyssetvalue{/bbe/j8/exceptions/Loop/SpecRequireCount}{0}
\pgfkeyssetvalue{/bbe/j8/exceptions/Loop/SpecEnsureCount}{0}
\pgfkeyssetvalue{/bbe/j8/exceptions/Loop/SpecRaiseCount}{4}
\pgfkeyssetvalue{/bbe/j8/exceptions/Loop/SpecReturnCount}{5}
\pgfkeyssetvalue{/bbe/j8/exceptions/Loop/SpecPredicateCount}{1}
\pgfkeyssetvalue{/bbe/j8/exceptions/Loop/SpecPureCount}{0}
\pgfkeyssetvalue{/bbe/j8/exceptions/Loop/SpecAssertionCount}{4}
\pgfkeyssetvalue{/bbe/j8/exceptions/Loop/SpecAssumptionCount}{0}
\pgfkeyssetvalue{/bbe/j8/exceptions/Loop/SpecInvariantCount}{3}
\pgfkeyssetvalue{/bbe/j8/exceptions/Loop/SpecExceptionCount}{9}
\pgfkeyssetvalue{/bbe/j8/exceptions/Loop/SpecFunctionalCount}{0}
\pgfkeyssetvalue{/bbe/j8/exceptions/Loop/SpecIntermediateCount}{7}
\pgfkeyssetvalue{/bbe/j8/exceptions/TryFinally/ConversionTime}{1017.2}
\pgfkeyssetvalue{/bbe/j8/exceptions/TryFinally/ConversionOverhead}{0.0711600449951428}
\pgfkeyssetvalue{/bbe/j8/exceptions/TryFinally/VerificationTime}{982.8}
\pgfkeyssetvalue{/bbe/j8/exceptions/TryFinally/SourceLinesOfCode}{125}
\pgfkeyssetvalue{/bbe/j8/exceptions/TryFinally/BytecodeLinesOfCode}{164}
\pgfkeyssetvalue{/bbe/j8/exceptions/TryFinally/BoogieLinesOfCode}{386}
\pgfkeyssetvalue{/bbe/j8/exceptions/TryFinally/MethodCount}{15}
\pgfkeyssetvalue{/bbe/j8/exceptions/TryFinally/SpecRequireCount}{0}
\pgfkeyssetvalue{/bbe/j8/exceptions/TryFinally/SpecEnsureCount}{4}
\pgfkeyssetvalue{/bbe/j8/exceptions/TryFinally/SpecRaiseCount}{2}
\pgfkeyssetvalue{/bbe/j8/exceptions/TryFinally/SpecReturnCount}{4}
\pgfkeyssetvalue{/bbe/j8/exceptions/TryFinally/SpecPredicateCount}{2}
\pgfkeyssetvalue{/bbe/j8/exceptions/TryFinally/SpecPureCount}{0}
\pgfkeyssetvalue{/bbe/j8/exceptions/TryFinally/SpecAssertionCount}{7}
\pgfkeyssetvalue{/bbe/j8/exceptions/TryFinally/SpecAssumptionCount}{0}
\pgfkeyssetvalue{/bbe/j8/exceptions/TryFinally/SpecInvariantCount}{0}
\pgfkeyssetvalue{/bbe/j8/exceptions/TryFinally/SpecExceptionCount}{6}
\pgfkeyssetvalue{/bbe/j8/exceptions/TryFinally/SpecFunctionalCount}{4}
\pgfkeyssetvalue{/bbe/j8/exceptions/TryFinally/SpecIntermediateCount}{7}
\pgfkeyssetvalue{/bbe/j8/exceptions/TryWithResources/ConversionTime}{1293.8}
\pgfkeyssetvalue{/bbe/j8/exceptions/TryWithResources/ConversionOverhead}{0.1520214657791404}
\pgfkeyssetvalue{/bbe/j8/exceptions/TryWithResources/VerificationTime}{1206.6}
\pgfkeyssetvalue{/bbe/j8/exceptions/TryWithResources/SourceLinesOfCode}{199}
\pgfkeyssetvalue{/bbe/j8/exceptions/TryWithResources/BytecodeLinesOfCode}{1584}
\pgfkeyssetvalue{/bbe/j8/exceptions/TryWithResources/BoogieLinesOfCode}{1635}
\pgfkeyssetvalue{/bbe/j8/exceptions/TryWithResources/MethodCount}{26}
\pgfkeyssetvalue{/bbe/j8/exceptions/TryWithResources/SpecRequireCount}{0}
\pgfkeyssetvalue{/bbe/j8/exceptions/TryWithResources/SpecEnsureCount}{4}
\pgfkeyssetvalue{/bbe/j8/exceptions/TryWithResources/SpecRaiseCount}{0}
\pgfkeyssetvalue{/bbe/j8/exceptions/TryWithResources/SpecReturnCount}{13}
\pgfkeyssetvalue{/bbe/j8/exceptions/TryWithResources/SpecPredicateCount}{3}
\pgfkeyssetvalue{/bbe/j8/exceptions/TryWithResources/SpecPureCount}{2}
\pgfkeyssetvalue{/bbe/j8/exceptions/TryWithResources/SpecAssertionCount}{39}
\pgfkeyssetvalue{/bbe/j8/exceptions/TryWithResources/SpecAssumptionCount}{0}
\pgfkeyssetvalue{/bbe/j8/exceptions/TryWithResources/SpecInvariantCount}{0}
\pgfkeyssetvalue{/bbe/j8/exceptions/TryWithResources/SpecExceptionCount}{13}
\pgfkeyssetvalue{/bbe/j8/exceptions/TryWithResources/SpecFunctionalCount}{4}
\pgfkeyssetvalue{/bbe/j8/exceptions/TryWithResources/SpecIntermediateCount}{39}
\pgfkeyssetvalue{/bbe/j8/algorithm/ArrayReverse/ConversionTime}{1118.4}
\pgfkeyssetvalue{/bbe/j8/algorithm/ArrayReverse/ConversionOverhead}{0.067294250574229}
\pgfkeyssetvalue{/bbe/j8/algorithm/ArrayReverse/VerificationTime}{1065.8}
\pgfkeyssetvalue{/bbe/j8/algorithm/ArrayReverse/SourceLinesOfCode}{72}
\pgfkeyssetvalue{/bbe/j8/algorithm/ArrayReverse/BytecodeLinesOfCode}{214}
\pgfkeyssetvalue{/bbe/j8/algorithm/ArrayReverse/BoogieLinesOfCode}{221}
\pgfkeyssetvalue{/bbe/j8/algorithm/ArrayReverse/MethodCount}{9}
\pgfkeyssetvalue{/bbe/j8/algorithm/ArrayReverse/SpecRequireCount}{1}
\pgfkeyssetvalue{/bbe/j8/algorithm/ArrayReverse/SpecEnsureCount}{2}
\pgfkeyssetvalue{/bbe/j8/algorithm/ArrayReverse/SpecRaiseCount}{1}
\pgfkeyssetvalue{/bbe/j8/algorithm/ArrayReverse/SpecReturnCount}{1}
\pgfkeyssetvalue{/bbe/j8/algorithm/ArrayReverse/SpecPredicateCount}{5}
\pgfkeyssetvalue{/bbe/j8/algorithm/ArrayReverse/SpecPureCount}{4}
\pgfkeyssetvalue{/bbe/j8/algorithm/ArrayReverse/SpecAssertionCount}{0}
\pgfkeyssetvalue{/bbe/j8/algorithm/ArrayReverse/SpecAssumptionCount}{0}
\pgfkeyssetvalue{/bbe/j8/algorithm/ArrayReverse/SpecInvariantCount}{3}
\pgfkeyssetvalue{/bbe/j8/algorithm/ArrayReverse/SpecExceptionCount}{2}
\pgfkeyssetvalue{/bbe/j8/algorithm/ArrayReverse/SpecFunctionalCount}{3}
\pgfkeyssetvalue{/bbe/j8/algorithm/ArrayReverse/SpecIntermediateCount}{3}
\pgfkeyssetvalue{/bbe/j8/algorithm/BinarySearch/ConversionTime}{1173.0}
\pgfkeyssetvalue{/bbe/j8/algorithm/BinarySearch/ConversionOverhead}{0.0717943317935084}
\pgfkeyssetvalue{/bbe/j8/algorithm/BinarySearch/VerificationTime}{819.0}
\pgfkeyssetvalue{/bbe/j8/algorithm/BinarySearch/SourceLinesOfCode}{52}
\pgfkeyssetvalue{/bbe/j8/algorithm/BinarySearch/BytecodeLinesOfCode}{124}
\pgfkeyssetvalue{/bbe/j8/algorithm/BinarySearch/BoogieLinesOfCode}{169}
\pgfkeyssetvalue{/bbe/j8/algorithm/BinarySearch/MethodCount}{6}
\pgfkeyssetvalue{/bbe/j8/algorithm/BinarySearch/SpecRequireCount}{3}
\pgfkeyssetvalue{/bbe/j8/algorithm/BinarySearch/SpecEnsureCount}{1}
\pgfkeyssetvalue{/bbe/j8/algorithm/BinarySearch/SpecRaiseCount}{0}
\pgfkeyssetvalue{/bbe/j8/algorithm/BinarySearch/SpecReturnCount}{1}
\pgfkeyssetvalue{/bbe/j8/algorithm/BinarySearch/SpecPredicateCount}{4}
\pgfkeyssetvalue{/bbe/j8/algorithm/BinarySearch/SpecPureCount}{0}
\pgfkeyssetvalue{/bbe/j8/algorithm/BinarySearch/SpecAssertionCount}{2}
\pgfkeyssetvalue{/bbe/j8/algorithm/BinarySearch/SpecAssumptionCount}{0}
\pgfkeyssetvalue{/bbe/j8/algorithm/BinarySearch/SpecInvariantCount}{0}
\pgfkeyssetvalue{/bbe/j8/algorithm/BinarySearch/SpecExceptionCount}{1}
\pgfkeyssetvalue{/bbe/j8/algorithm/BinarySearch/SpecFunctionalCount}{4}
\pgfkeyssetvalue{/bbe/j8/algorithm/BinarySearch/SpecIntermediateCount}{2}
\pgfkeyssetvalue{/bbe/j8/algorithm/GCD/ConversionTime}{1076.8}
\pgfkeyssetvalue{/bbe/j8/algorithm/GCD/ConversionOverhead}{0.0632467903525743}
\pgfkeyssetvalue{/bbe/j8/algorithm/GCD/VerificationTime}{786.0}
\pgfkeyssetvalue{/bbe/j8/algorithm/GCD/SourceLinesOfCode}{50}
\pgfkeyssetvalue{/bbe/j8/algorithm/GCD/BytecodeLinesOfCode}{112}
\pgfkeyssetvalue{/bbe/j8/algorithm/GCD/BoogieLinesOfCode}{200}
\pgfkeyssetvalue{/bbe/j8/algorithm/GCD/MethodCount}{6}
\pgfkeyssetvalue{/bbe/j8/algorithm/GCD/SpecRequireCount}{0}
\pgfkeyssetvalue{/bbe/j8/algorithm/GCD/SpecEnsureCount}{1}
\pgfkeyssetvalue{/bbe/j8/algorithm/GCD/SpecRaiseCount}{1}
\pgfkeyssetvalue{/bbe/j8/algorithm/GCD/SpecReturnCount}{0}
\pgfkeyssetvalue{/bbe/j8/algorithm/GCD/SpecPredicateCount}{3}
\pgfkeyssetvalue{/bbe/j8/algorithm/GCD/SpecPureCount}{3}
\pgfkeyssetvalue{/bbe/j8/algorithm/GCD/SpecAssertionCount}{0}
\pgfkeyssetvalue{/bbe/j8/algorithm/GCD/SpecAssumptionCount}{0}
\pgfkeyssetvalue{/bbe/j8/algorithm/GCD/SpecInvariantCount}{2}
\pgfkeyssetvalue{/bbe/j8/algorithm/GCD/SpecExceptionCount}{1}
\pgfkeyssetvalue{/bbe/j8/algorithm/GCD/SpecFunctionalCount}{1}
\pgfkeyssetvalue{/bbe/j8/algorithm/GCD/SpecIntermediateCount}{2}
\pgfkeyssetvalue{/bbe/j8/algorithm/LinearSearch/ConversionTime}{1152.0}
\pgfkeyssetvalue{/bbe/j8/algorithm/LinearSearch/ConversionOverhead}{0.0687173260466327}
\pgfkeyssetvalue{/bbe/j8/algorithm/LinearSearch/VerificationTime}{812.2}
\pgfkeyssetvalue{/bbe/j8/algorithm/LinearSearch/SourceLinesOfCode}{62}
\pgfkeyssetvalue{/bbe/j8/algorithm/LinearSearch/BytecodeLinesOfCode}{126}
\pgfkeyssetvalue{/bbe/j8/algorithm/LinearSearch/BoogieLinesOfCode}{193}
\pgfkeyssetvalue{/bbe/j8/algorithm/LinearSearch/MethodCount}{9}
\pgfkeyssetvalue{/bbe/j8/algorithm/LinearSearch/SpecRequireCount}{4}
\pgfkeyssetvalue{/bbe/j8/algorithm/LinearSearch/SpecEnsureCount}{2}
\pgfkeyssetvalue{/bbe/j8/algorithm/LinearSearch/SpecRaiseCount}{0}
\pgfkeyssetvalue{/bbe/j8/algorithm/LinearSearch/SpecReturnCount}{2}
\pgfkeyssetvalue{/bbe/j8/algorithm/LinearSearch/SpecPredicateCount}{6}
\pgfkeyssetvalue{/bbe/j8/algorithm/LinearSearch/SpecPureCount}{0}
\pgfkeyssetvalue{/bbe/j8/algorithm/LinearSearch/SpecAssertionCount}{0}
\pgfkeyssetvalue{/bbe/j8/algorithm/LinearSearch/SpecAssumptionCount}{0}
\pgfkeyssetvalue{/bbe/j8/algorithm/LinearSearch/SpecInvariantCount}{2}
\pgfkeyssetvalue{/bbe/j8/algorithm/LinearSearch/SpecExceptionCount}{2}
\pgfkeyssetvalue{/bbe/j8/algorithm/LinearSearch/SpecFunctionalCount}{6}
\pgfkeyssetvalue{/bbe/j8/algorithm/LinearSearch/SpecIntermediateCount}{2}
\pgfkeyssetvalue{/bbe/j8/algorithm/DoubleSelectionSort/ConversionTime}{1217.4}
\pgfkeyssetvalue{/bbe/j8/algorithm/DoubleSelectionSort/ConversionOverhead}{0.0728774100157692}
\pgfkeyssetvalue{/bbe/j8/algorithm/DoubleSelectionSort/VerificationTime}{1895.2}
\pgfkeyssetvalue{/bbe/j8/algorithm/DoubleSelectionSort/SourceLinesOfCode}{110}
\pgfkeyssetvalue{/bbe/j8/algorithm/DoubleSelectionSort/BytecodeLinesOfCode}{296}
\pgfkeyssetvalue{/bbe/j8/algorithm/DoubleSelectionSort/BoogieLinesOfCode}{234}
\pgfkeyssetvalue{/bbe/j8/algorithm/DoubleSelectionSort/MethodCount}{16}
\pgfkeyssetvalue{/bbe/j8/algorithm/DoubleSelectionSort/SpecRequireCount}{4}
\pgfkeyssetvalue{/bbe/j8/algorithm/DoubleSelectionSort/SpecEnsureCount}{5}
\pgfkeyssetvalue{/bbe/j8/algorithm/DoubleSelectionSort/SpecRaiseCount}{0}
\pgfkeyssetvalue{/bbe/j8/algorithm/DoubleSelectionSort/SpecReturnCount}{3}
\pgfkeyssetvalue{/bbe/j8/algorithm/DoubleSelectionSort/SpecPredicateCount}{9}
\pgfkeyssetvalue{/bbe/j8/algorithm/DoubleSelectionSort/SpecPureCount}{6}
\pgfkeyssetvalue{/bbe/j8/algorithm/DoubleSelectionSort/SpecAssertionCount}{0}
\pgfkeyssetvalue{/bbe/j8/algorithm/DoubleSelectionSort/SpecAssumptionCount}{0}
\pgfkeyssetvalue{/bbe/j8/algorithm/DoubleSelectionSort/SpecInvariantCount}{6}
\pgfkeyssetvalue{/bbe/j8/algorithm/DoubleSelectionSort/SpecExceptionCount}{3}
\pgfkeyssetvalue{/bbe/j8/algorithm/DoubleSelectionSort/SpecFunctionalCount}{9}
\pgfkeyssetvalue{/bbe/j8/algorithm/DoubleSelectionSort/SpecIntermediateCount}{6}
\pgfkeyssetvalue{/bbe/j8/algorithm/IntegerSelectionSort/ConversionTime}{1134.8}
\pgfkeyssetvalue{/bbe/j8/algorithm/IntegerSelectionSort/ConversionOverhead}{0.0701530581266474}
\pgfkeyssetvalue{/bbe/j8/algorithm/IntegerSelectionSort/VerificationTime}{3210.8}
\pgfkeyssetvalue{/bbe/j8/algorithm/IntegerSelectionSort/SourceLinesOfCode}{110}
\pgfkeyssetvalue{/bbe/j8/algorithm/IntegerSelectionSort/BytecodeLinesOfCode}{295}
\pgfkeyssetvalue{/bbe/j8/algorithm/IntegerSelectionSort/BoogieLinesOfCode}{234}
\pgfkeyssetvalue{/bbe/j8/algorithm/IntegerSelectionSort/MethodCount}{16}
\pgfkeyssetvalue{/bbe/j8/algorithm/IntegerSelectionSort/SpecRequireCount}{4}
\pgfkeyssetvalue{/bbe/j8/algorithm/IntegerSelectionSort/SpecEnsureCount}{5}
\pgfkeyssetvalue{/bbe/j8/algorithm/IntegerSelectionSort/SpecRaiseCount}{0}
\pgfkeyssetvalue{/bbe/j8/algorithm/IntegerSelectionSort/SpecReturnCount}{3}
\pgfkeyssetvalue{/bbe/j8/algorithm/IntegerSelectionSort/SpecPredicateCount}{9}
\pgfkeyssetvalue{/bbe/j8/algorithm/IntegerSelectionSort/SpecPureCount}{6}
\pgfkeyssetvalue{/bbe/j8/algorithm/IntegerSelectionSort/SpecAssertionCount}{0}
\pgfkeyssetvalue{/bbe/j8/algorithm/IntegerSelectionSort/SpecAssumptionCount}{0}
\pgfkeyssetvalue{/bbe/j8/algorithm/IntegerSelectionSort/SpecInvariantCount}{6}
\pgfkeyssetvalue{/bbe/j8/algorithm/IntegerSelectionSort/SpecExceptionCount}{3}
\pgfkeyssetvalue{/bbe/j8/algorithm/IntegerSelectionSort/SpecFunctionalCount}{9}
\pgfkeyssetvalue{/bbe/j8/algorithm/IntegerSelectionSort/SpecIntermediateCount}{6}
\pgfkeyssetvalue{/bbe/j8/algorithm/SquareSortedArray/ConversionTime}{1113.6}
\pgfkeyssetvalue{/bbe/j8/algorithm/SquareSortedArray/ConversionOverhead}{0.0601136833611625}
\pgfkeyssetvalue{/bbe/j8/algorithm/SquareSortedArray/VerificationTime}{790.2}
\pgfkeyssetvalue{/bbe/j8/algorithm/SquareSortedArray/SourceLinesOfCode}{61}
\pgfkeyssetvalue{/bbe/j8/algorithm/SquareSortedArray/BytecodeLinesOfCode}{135}
\pgfkeyssetvalue{/bbe/j8/algorithm/SquareSortedArray/BoogieLinesOfCode}{187}
\pgfkeyssetvalue{/bbe/j8/algorithm/SquareSortedArray/MethodCount}{7}
\pgfkeyssetvalue{/bbe/j8/algorithm/SquareSortedArray/SpecRequireCount}{1}
\pgfkeyssetvalue{/bbe/j8/algorithm/SquareSortedArray/SpecEnsureCount}{0}
\pgfkeyssetvalue{/bbe/j8/algorithm/SquareSortedArray/SpecRaiseCount}{1}
\pgfkeyssetvalue{/bbe/j8/algorithm/SquareSortedArray/SpecReturnCount}{0}
\pgfkeyssetvalue{/bbe/j8/algorithm/SquareSortedArray/SpecPredicateCount}{4}
\pgfkeyssetvalue{/bbe/j8/algorithm/SquareSortedArray/SpecPureCount}{1}
\pgfkeyssetvalue{/bbe/j8/algorithm/SquareSortedArray/SpecAssertionCount}{0}
\pgfkeyssetvalue{/bbe/j8/algorithm/SquareSortedArray/SpecAssumptionCount}{0}
\pgfkeyssetvalue{/bbe/j8/algorithm/SquareSortedArray/SpecInvariantCount}{2}
\pgfkeyssetvalue{/bbe/j8/algorithm/SquareSortedArray/SpecExceptionCount}{1}
\pgfkeyssetvalue{/bbe/j8/algorithm/SquareSortedArray/SpecFunctionalCount}{1}
\pgfkeyssetvalue{/bbe/j8/algorithm/SquareSortedArray/SpecIntermediateCount}{2}
\pgfkeyssetvalue{/bbe/j8/algorithm/IntegerSum/ConversionTime}{1094.6}
\pgfkeyssetvalue{/bbe/j8/algorithm/IntegerSum/ConversionOverhead}{0.061791354562425}
\pgfkeyssetvalue{/bbe/j8/algorithm/IntegerSum/VerificationTime}{774.4}
\pgfkeyssetvalue{/bbe/j8/algorithm/IntegerSum/SourceLinesOfCode}{45}
\pgfkeyssetvalue{/bbe/j8/algorithm/IntegerSum/BytecodeLinesOfCode}{86}
\pgfkeyssetvalue{/bbe/j8/algorithm/IntegerSum/BoogieLinesOfCode}{175}
\pgfkeyssetvalue{/bbe/j8/algorithm/IntegerSum/MethodCount}{5}
\pgfkeyssetvalue{/bbe/j8/algorithm/IntegerSum/SpecRequireCount}{0}
\pgfkeyssetvalue{/bbe/j8/algorithm/IntegerSum/SpecEnsureCount}{1}
\pgfkeyssetvalue{/bbe/j8/algorithm/IntegerSum/SpecRaiseCount}{1}
\pgfkeyssetvalue{/bbe/j8/algorithm/IntegerSum/SpecReturnCount}{0}
\pgfkeyssetvalue{/bbe/j8/algorithm/IntegerSum/SpecPredicateCount}{2}
\pgfkeyssetvalue{/bbe/j8/algorithm/IntegerSum/SpecPureCount}{3}
\pgfkeyssetvalue{/bbe/j8/algorithm/IntegerSum/SpecAssertionCount}{0}
\pgfkeyssetvalue{/bbe/j8/algorithm/IntegerSum/SpecAssumptionCount}{0}
\pgfkeyssetvalue{/bbe/j8/algorithm/IntegerSum/SpecInvariantCount}{3}
\pgfkeyssetvalue{/bbe/j8/algorithm/IntegerSum/SpecExceptionCount}{1}
\pgfkeyssetvalue{/bbe/j8/algorithm/IntegerSum/SpecFunctionalCount}{1}
\pgfkeyssetvalue{/bbe/j8/algorithm/IntegerSum/SpecIntermediateCount}{3}
\pgfkeyssetvalue{/bbe/j8/library/java/util/ArrayList/ConversionTime}{5714.8}
\pgfkeyssetvalue{/bbe/j8/library/java/util/ArrayList/ConversionOverhead}{0.7881850667270107}
\pgfkeyssetvalue{/bbe/j8/library/java/util/ArrayList/VerificationTime}{5701.2}
\pgfkeyssetvalue{/bbe/j8/library/java/util/ArrayList/SourceLinesOfCode}{2653}
\pgfkeyssetvalue{/bbe/j8/library/java/util/ArrayList/BytecodeLinesOfCode}{6999}
\pgfkeyssetvalue{/bbe/j8/library/java/util/ArrayList/BoogieLinesOfCode}{7160}
\pgfkeyssetvalue{/bbe/j8/library/java/util/ArrayList/MethodCount}{294}
\pgfkeyssetvalue{/bbe/j8/library/java/util/ArrayList/SpecRequireCount}{0}
\pgfkeyssetvalue{/bbe/j8/library/java/util/ArrayList/SpecEnsureCount}{0}
\pgfkeyssetvalue{/bbe/j8/library/java/util/ArrayList/SpecRaiseCount}{19}
\pgfkeyssetvalue{/bbe/j8/library/java/util/ArrayList/SpecReturnCount}{5}
\pgfkeyssetvalue{/bbe/j8/library/java/util/ArrayList/SpecPredicateCount}{14}
\pgfkeyssetvalue{/bbe/j8/library/java/util/ArrayList/SpecPureCount}{8}
\pgfkeyssetvalue{/bbe/j8/library/java/util/ArrayList/SpecAssertionCount}{0}
\pgfkeyssetvalue{/bbe/j8/library/java/util/ArrayList/SpecAssumptionCount}{0}
\pgfkeyssetvalue{/bbe/j8/library/java/util/ArrayList/SpecInvariantCount}{0}
\pgfkeyssetvalue{/bbe/j8/library/java/util/ArrayList/SpecExceptionCount}{24}
\pgfkeyssetvalue{/bbe/j8/library/java/util/ArrayList/SpecFunctionalCount}{0}
\pgfkeyssetvalue{/bbe/j8/library/java/util/ArrayList/SpecIntermediateCount}{0}
\pgfkeyssetvalue{/bbe/j8/library/java/util/LinkedList/ConversionTime}{2083.6}
\pgfkeyssetvalue{/bbe/j8/library/java/util/LinkedList/ConversionOverhead}{0.4985376493694324}
\pgfkeyssetvalue{/bbe/j8/library/java/util/LinkedList/VerificationTime}{2902.8}
\pgfkeyssetvalue{/bbe/j8/library/java/util/LinkedList/SourceLinesOfCode}{2472}
\pgfkeyssetvalue{/bbe/j8/library/java/util/LinkedList/BytecodeLinesOfCode}{6271}
\pgfkeyssetvalue{/bbe/j8/library/java/util/LinkedList/BoogieLinesOfCode}{3041}
\pgfkeyssetvalue{/bbe/j8/library/java/util/LinkedList/MethodCount}{366}
\pgfkeyssetvalue{/bbe/j8/library/java/util/LinkedList/SpecRequireCount}{0}
\pgfkeyssetvalue{/bbe/j8/library/java/util/LinkedList/SpecEnsureCount}{2}
\pgfkeyssetvalue{/bbe/j8/library/java/util/LinkedList/SpecRaiseCount}{10}
\pgfkeyssetvalue{/bbe/j8/library/java/util/LinkedList/SpecReturnCount}{7}
\pgfkeyssetvalue{/bbe/j8/library/java/util/LinkedList/SpecPredicateCount}{8}
\pgfkeyssetvalue{/bbe/j8/library/java/util/LinkedList/SpecPureCount}{0}
\pgfkeyssetvalue{/bbe/j8/library/java/util/LinkedList/SpecAssertionCount}{0}
\pgfkeyssetvalue{/bbe/j8/library/java/util/LinkedList/SpecAssumptionCount}{0}
\pgfkeyssetvalue{/bbe/j8/library/java/util/LinkedList/SpecInvariantCount}{0}
\pgfkeyssetvalue{/bbe/j8/library/java/util/LinkedList/SpecExceptionCount}{17}
\pgfkeyssetvalue{/bbe/j8/library/java/util/LinkedList/SpecFunctionalCount}{2}
\pgfkeyssetvalue{/bbe/j8/library/java/util/LinkedList/SpecIntermediateCount}{0}
\pgfkeyssetvalue{/bbe/j17/exceptions/TryWithResources/ConversionTime}{1013.2}
\pgfkeyssetvalue{/bbe/j17/exceptions/TryWithResources/ConversionOverhead}{0.0611347224858567}
\pgfkeyssetvalue{/bbe/j17/exceptions/TryWithResources/VerificationTime}{856.0}
\pgfkeyssetvalue{/bbe/j17/exceptions/TryWithResources/SourceLinesOfCode}{44}
\pgfkeyssetvalue{/bbe/j17/exceptions/TryWithResources/BytecodeLinesOfCode}{104}
\pgfkeyssetvalue{/bbe/j17/exceptions/TryWithResources/BoogieLinesOfCode}{220}
\pgfkeyssetvalue{/bbe/j17/exceptions/TryWithResources/MethodCount}{6}
\pgfkeyssetvalue{/bbe/j17/exceptions/TryWithResources/SpecRequireCount}{0}
\pgfkeyssetvalue{/bbe/j17/exceptions/TryWithResources/SpecEnsureCount}{1}
\pgfkeyssetvalue{/bbe/j17/exceptions/TryWithResources/SpecRaiseCount}{0}
\pgfkeyssetvalue{/bbe/j17/exceptions/TryWithResources/SpecReturnCount}{1}
\pgfkeyssetvalue{/bbe/j17/exceptions/TryWithResources/SpecPredicateCount}{1}
\pgfkeyssetvalue{/bbe/j17/exceptions/TryWithResources/SpecPureCount}{1}
\pgfkeyssetvalue{/bbe/j17/exceptions/TryWithResources/SpecAssertionCount}{2}
\pgfkeyssetvalue{/bbe/j17/exceptions/TryWithResources/SpecAssumptionCount}{0}
\pgfkeyssetvalue{/bbe/j17/exceptions/TryWithResources/SpecInvariantCount}{0}
\pgfkeyssetvalue{/bbe/j17/exceptions/TryWithResources/SpecExceptionCount}{1}
\pgfkeyssetvalue{/bbe/j17/exceptions/TryWithResources/SpecFunctionalCount}{1}
\pgfkeyssetvalue{/bbe/j17/exceptions/TryWithResources/SpecIntermediateCount}{2}
\pgfkeyssetvalue{/bbe/j17/examples/Summary/ConversionTime}{927.0}
\pgfkeyssetvalue{/bbe/j17/examples/Summary/ConversionOverhead}{0.0471181253054825}
\pgfkeyssetvalue{/bbe/j17/examples/Summary/VerificationTime}{809.0}
\pgfkeyssetvalue{/bbe/j17/examples/Summary/SourceLinesOfCode}{48}
\pgfkeyssetvalue{/bbe/j17/examples/Summary/BytecodeLinesOfCode}{88}
\pgfkeyssetvalue{/bbe/j17/examples/Summary/BoogieLinesOfCode}{171}
\pgfkeyssetvalue{/bbe/j17/examples/Summary/MethodCount}{5}
\pgfkeyssetvalue{/bbe/j17/examples/Summary/SpecRequireCount}{1}
\pgfkeyssetvalue{/bbe/j17/examples/Summary/SpecEnsureCount}{1}
\pgfkeyssetvalue{/bbe/j17/examples/Summary/SpecRaiseCount}{0}
\pgfkeyssetvalue{/bbe/j17/examples/Summary/SpecReturnCount}{1}
\pgfkeyssetvalue{/bbe/j17/examples/Summary/SpecPredicateCount}{2}
\pgfkeyssetvalue{/bbe/j17/examples/Summary/SpecPureCount}{1}
\pgfkeyssetvalue{/bbe/j17/examples/Summary/SpecAssertionCount}{0}
\pgfkeyssetvalue{/bbe/j17/examples/Summary/SpecAssumptionCount}{0}
\pgfkeyssetvalue{/bbe/j17/examples/Summary/SpecInvariantCount}{1}
\pgfkeyssetvalue{/bbe/j17/examples/Summary/SpecExceptionCount}{1}
\pgfkeyssetvalue{/bbe/j17/examples/Summary/SpecFunctionalCount}{2}
\pgfkeyssetvalue{/bbe/j17/examples/Summary/SpecIntermediateCount}{1}
\pgfkeyssetvalue{/bbe/j17/examples/ReadResource/ConversionTime}{1085.6}
\pgfkeyssetvalue{/bbe/j17/examples/ReadResource/ConversionOverhead}{0.069865182895389}
\pgfkeyssetvalue{/bbe/j17/examples/ReadResource/VerificationTime}{843.8}
\pgfkeyssetvalue{/bbe/j17/examples/ReadResource/SourceLinesOfCode}{117}
\pgfkeyssetvalue{/bbe/j17/examples/ReadResource/BytecodeLinesOfCode}{182}
\pgfkeyssetvalue{/bbe/j17/examples/ReadResource/BoogieLinesOfCode}{447}
\pgfkeyssetvalue{/bbe/j17/examples/ReadResource/MethodCount}{14}
\pgfkeyssetvalue{/bbe/j17/examples/ReadResource/SpecRequireCount}{1}
\pgfkeyssetvalue{/bbe/j17/examples/ReadResource/SpecEnsureCount}{5}
\pgfkeyssetvalue{/bbe/j17/examples/ReadResource/SpecRaiseCount}{4}
\pgfkeyssetvalue{/bbe/j17/examples/ReadResource/SpecReturnCount}{3}
\pgfkeyssetvalue{/bbe/j17/examples/ReadResource/SpecPredicateCount}{12}
\pgfkeyssetvalue{/bbe/j17/examples/ReadResource/SpecPureCount}{9}
\pgfkeyssetvalue{/bbe/j17/examples/ReadResource/SpecAssertionCount}{0}
\pgfkeyssetvalue{/bbe/j17/examples/ReadResource/SpecAssumptionCount}{0}
\pgfkeyssetvalue{/bbe/j17/examples/ReadResource/SpecInvariantCount}{1}
\pgfkeyssetvalue{/bbe/j17/examples/ReadResource/SpecExceptionCount}{7}
\pgfkeyssetvalue{/bbe/j17/examples/ReadResource/SpecFunctionalCount}{6}
\pgfkeyssetvalue{/bbe/j17/examples/ReadResource/SpecIntermediateCount}{1}
\pgfkeyssetvalue{/bbe/s2/exceptions/PotentialIndexOutOfBounds/ConversionTime}{1212.8}
\pgfkeyssetvalue{/bbe/s2/exceptions/PotentialIndexOutOfBounds/ConversionOverhead}{0.0683337834678026}
\pgfkeyssetvalue{/bbe/s2/exceptions/PotentialIndexOutOfBounds/VerificationTime}{826.8}
\pgfkeyssetvalue{/bbe/s2/exceptions/PotentialIndexOutOfBounds/SourceLinesOfCode}{44}
\pgfkeyssetvalue{/bbe/s2/exceptions/PotentialIndexOutOfBounds/BytecodeLinesOfCode}{65}
\pgfkeyssetvalue{/bbe/s2/exceptions/PotentialIndexOutOfBounds/BoogieLinesOfCode}{231}
\pgfkeyssetvalue{/bbe/s2/exceptions/PotentialIndexOutOfBounds/MethodCount}{7}
\pgfkeyssetvalue{/bbe/s2/exceptions/PotentialIndexOutOfBounds/SpecRequireCount}{1}
\pgfkeyssetvalue{/bbe/s2/exceptions/PotentialIndexOutOfBounds/SpecEnsureCount}{0}
\pgfkeyssetvalue{/bbe/s2/exceptions/PotentialIndexOutOfBounds/SpecRaiseCount}{1}
\pgfkeyssetvalue{/bbe/s2/exceptions/PotentialIndexOutOfBounds/SpecReturnCount}{3}
\pgfkeyssetvalue{/bbe/s2/exceptions/PotentialIndexOutOfBounds/SpecPredicateCount}{2}
\pgfkeyssetvalue{/bbe/s2/exceptions/PotentialIndexOutOfBounds/SpecPureCount}{0}
\pgfkeyssetvalue{/bbe/s2/exceptions/PotentialIndexOutOfBounds/SpecAssertionCount}{0}
\pgfkeyssetvalue{/bbe/s2/exceptions/PotentialIndexOutOfBounds/SpecAssumptionCount}{0}
\pgfkeyssetvalue{/bbe/s2/exceptions/PotentialIndexOutOfBounds/SpecInvariantCount}{0}
\pgfkeyssetvalue{/bbe/s2/exceptions/PotentialIndexOutOfBounds/SpecExceptionCount}{4}
\pgfkeyssetvalue{/bbe/s2/exceptions/PotentialIndexOutOfBounds/SpecFunctionalCount}{1}
\pgfkeyssetvalue{/bbe/s2/exceptions/PotentialIndexOutOfBounds/SpecIntermediateCount}{0}
\pgfkeyssetvalue{/bbe/s2/exceptions/PotentialNullDereference/ConversionTime}{1045.0}
\pgfkeyssetvalue{/bbe/s2/exceptions/PotentialNullDereference/ConversionOverhead}{0.071204934093064}
\pgfkeyssetvalue{/bbe/s2/exceptions/PotentialNullDereference/VerificationTime}{923.0}
\pgfkeyssetvalue{/bbe/s2/exceptions/PotentialNullDereference/SourceLinesOfCode}{43}
\pgfkeyssetvalue{/bbe/s2/exceptions/PotentialNullDereference/BytecodeLinesOfCode}{48}
\pgfkeyssetvalue{/bbe/s2/exceptions/PotentialNullDereference/BoogieLinesOfCode}{275}
\pgfkeyssetvalue{/bbe/s2/exceptions/PotentialNullDereference/MethodCount}{6}
\pgfkeyssetvalue{/bbe/s2/exceptions/PotentialNullDereference/SpecRequireCount}{0}
\pgfkeyssetvalue{/bbe/s2/exceptions/PotentialNullDereference/SpecEnsureCount}{0}
\pgfkeyssetvalue{/bbe/s2/exceptions/PotentialNullDereference/SpecRaiseCount}{0}
\pgfkeyssetvalue{/bbe/s2/exceptions/PotentialNullDereference/SpecReturnCount}{6}
\pgfkeyssetvalue{/bbe/s2/exceptions/PotentialNullDereference/SpecPredicateCount}{1}
\pgfkeyssetvalue{/bbe/s2/exceptions/PotentialNullDereference/SpecPureCount}{0}
\pgfkeyssetvalue{/bbe/s2/exceptions/PotentialNullDereference/SpecAssertionCount}{0}
\pgfkeyssetvalue{/bbe/s2/exceptions/PotentialNullDereference/SpecAssumptionCount}{0}
\pgfkeyssetvalue{/bbe/s2/exceptions/PotentialNullDereference/SpecInvariantCount}{0}
\pgfkeyssetvalue{/bbe/s2/exceptions/PotentialNullDereference/SpecExceptionCount}{6}
\pgfkeyssetvalue{/bbe/s2/exceptions/PotentialNullDereference/SpecFunctionalCount}{0}
\pgfkeyssetvalue{/bbe/s2/exceptions/PotentialNullDereference/SpecIntermediateCount}{0}
\pgfkeyssetvalue{/bbe/s2/exceptions/MultiCatch/ConversionTime}{1173.6}
\pgfkeyssetvalue{/bbe/s2/exceptions/MultiCatch/ConversionOverhead}{0.067787989814557}
\pgfkeyssetvalue{/bbe/s2/exceptions/MultiCatch/VerificationTime}{858.6}
\pgfkeyssetvalue{/bbe/s2/exceptions/MultiCatch/SourceLinesOfCode}{45}
\pgfkeyssetvalue{/bbe/s2/exceptions/MultiCatch/BytecodeLinesOfCode}{122}
\pgfkeyssetvalue{/bbe/s2/exceptions/MultiCatch/BoogieLinesOfCode}{297}
\pgfkeyssetvalue{/bbe/s2/exceptions/MultiCatch/MethodCount}{7}
\pgfkeyssetvalue{/bbe/s2/exceptions/MultiCatch/SpecRequireCount}{0}
\pgfkeyssetvalue{/bbe/s2/exceptions/MultiCatch/SpecEnsureCount}{1}
\pgfkeyssetvalue{/bbe/s2/exceptions/MultiCatch/SpecRaiseCount}{0}
\pgfkeyssetvalue{/bbe/s2/exceptions/MultiCatch/SpecReturnCount}{2}
\pgfkeyssetvalue{/bbe/s2/exceptions/MultiCatch/SpecPredicateCount}{1}
\pgfkeyssetvalue{/bbe/s2/exceptions/MultiCatch/SpecPureCount}{0}
\pgfkeyssetvalue{/bbe/s2/exceptions/MultiCatch/SpecAssertionCount}{2}
\pgfkeyssetvalue{/bbe/s2/exceptions/MultiCatch/SpecAssumptionCount}{0}
\pgfkeyssetvalue{/bbe/s2/exceptions/MultiCatch/SpecInvariantCount}{0}
\pgfkeyssetvalue{/bbe/s2/exceptions/MultiCatch/SpecExceptionCount}{2}
\pgfkeyssetvalue{/bbe/s2/exceptions/MultiCatch/SpecFunctionalCount}{1}
\pgfkeyssetvalue{/bbe/s2/exceptions/MultiCatch/SpecIntermediateCount}{2}
\pgfkeyssetvalue{/bbe/s2/exceptions/Basic/ConversionTime}{1142.6}
\pgfkeyssetvalue{/bbe/s2/exceptions/Basic/ConversionOverhead}{0.0935149676741878}
\pgfkeyssetvalue{/bbe/s2/exceptions/Basic/VerificationTime}{1084.8}
\pgfkeyssetvalue{/bbe/s2/exceptions/Basic/SourceLinesOfCode}{121}
\pgfkeyssetvalue{/bbe/s2/exceptions/Basic/BytecodeLinesOfCode}{207}
\pgfkeyssetvalue{/bbe/s2/exceptions/Basic/BoogieLinesOfCode}{460}
\pgfkeyssetvalue{/bbe/s2/exceptions/Basic/MethodCount}{22}
\pgfkeyssetvalue{/bbe/s2/exceptions/Basic/SpecRequireCount}{0}
\pgfkeyssetvalue{/bbe/s2/exceptions/Basic/SpecEnsureCount}{1}
\pgfkeyssetvalue{/bbe/s2/exceptions/Basic/SpecRaiseCount}{6}
\pgfkeyssetvalue{/bbe/s2/exceptions/Basic/SpecReturnCount}{6}
\pgfkeyssetvalue{/bbe/s2/exceptions/Basic/SpecPredicateCount}{7}
\pgfkeyssetvalue{/bbe/s2/exceptions/Basic/SpecPureCount}{6}
\pgfkeyssetvalue{/bbe/s2/exceptions/Basic/SpecAssertionCount}{4}
\pgfkeyssetvalue{/bbe/s2/exceptions/Basic/SpecAssumptionCount}{0}
\pgfkeyssetvalue{/bbe/s2/exceptions/Basic/SpecInvariantCount}{0}
\pgfkeyssetvalue{/bbe/s2/exceptions/Basic/SpecExceptionCount}{12}
\pgfkeyssetvalue{/bbe/s2/exceptions/Basic/SpecFunctionalCount}{1}
\pgfkeyssetvalue{/bbe/s2/exceptions/Basic/SpecIntermediateCount}{4}
\pgfkeyssetvalue{/bbe/s2/exceptions/TryFinally/ConversionTime}{1266.8}
\pgfkeyssetvalue{/bbe/s2/exceptions/TryFinally/ConversionOverhead}{0.0936041950165302}
\pgfkeyssetvalue{/bbe/s2/exceptions/TryFinally/VerificationTime}{1043.6}
\pgfkeyssetvalue{/bbe/s2/exceptions/TryFinally/SourceLinesOfCode}{117}
\pgfkeyssetvalue{/bbe/s2/exceptions/TryFinally/BytecodeLinesOfCode}{175}
\pgfkeyssetvalue{/bbe/s2/exceptions/TryFinally/BoogieLinesOfCode}{455}
\pgfkeyssetvalue{/bbe/s2/exceptions/TryFinally/MethodCount}{15}
\pgfkeyssetvalue{/bbe/s2/exceptions/TryFinally/SpecRequireCount}{0}
\pgfkeyssetvalue{/bbe/s2/exceptions/TryFinally/SpecEnsureCount}{4}
\pgfkeyssetvalue{/bbe/s2/exceptions/TryFinally/SpecRaiseCount}{1}
\pgfkeyssetvalue{/bbe/s2/exceptions/TryFinally/SpecReturnCount}{2}
\pgfkeyssetvalue{/bbe/s2/exceptions/TryFinally/SpecPredicateCount}{2}
\pgfkeyssetvalue{/bbe/s2/exceptions/TryFinally/SpecPureCount}{0}
\pgfkeyssetvalue{/bbe/s2/exceptions/TryFinally/SpecAssertionCount}{7}
\pgfkeyssetvalue{/bbe/s2/exceptions/TryFinally/SpecAssumptionCount}{0}
\pgfkeyssetvalue{/bbe/s2/exceptions/TryFinally/SpecInvariantCount}{0}
\pgfkeyssetvalue{/bbe/s2/exceptions/TryFinally/SpecExceptionCount}{3}
\pgfkeyssetvalue{/bbe/s2/exceptions/TryFinally/SpecFunctionalCount}{4}
\pgfkeyssetvalue{/bbe/s2/exceptions/TryFinally/SpecIntermediateCount}{7}
\pgfkeyssetvalue{/bbe/s2/algorithm/ArrayReverse/ConversionTime}{1117.4}
\pgfkeyssetvalue{/bbe/s2/algorithm/ArrayReverse/ConversionOverhead}{0.0710554370598876}
\pgfkeyssetvalue{/bbe/s2/algorithm/ArrayReverse/VerificationTime}{1214.0}
\pgfkeyssetvalue{/bbe/s2/algorithm/ArrayReverse/SourceLinesOfCode}{62}
\pgfkeyssetvalue{/bbe/s2/algorithm/ArrayReverse/BytecodeLinesOfCode}{207}
\pgfkeyssetvalue{/bbe/s2/algorithm/ArrayReverse/BoogieLinesOfCode}{221}
\pgfkeyssetvalue{/bbe/s2/algorithm/ArrayReverse/MethodCount}{8}
\pgfkeyssetvalue{/bbe/s2/algorithm/ArrayReverse/SpecRequireCount}{0}
\pgfkeyssetvalue{/bbe/s2/algorithm/ArrayReverse/SpecEnsureCount}{2}
\pgfkeyssetvalue{/bbe/s2/algorithm/ArrayReverse/SpecRaiseCount}{1}
\pgfkeyssetvalue{/bbe/s2/algorithm/ArrayReverse/SpecReturnCount}{1}
\pgfkeyssetvalue{/bbe/s2/algorithm/ArrayReverse/SpecPredicateCount}{4}
\pgfkeyssetvalue{/bbe/s2/algorithm/ArrayReverse/SpecPureCount}{2}
\pgfkeyssetvalue{/bbe/s2/algorithm/ArrayReverse/SpecAssertionCount}{0}
\pgfkeyssetvalue{/bbe/s2/algorithm/ArrayReverse/SpecAssumptionCount}{0}
\pgfkeyssetvalue{/bbe/s2/algorithm/ArrayReverse/SpecInvariantCount}{3}
\pgfkeyssetvalue{/bbe/s2/algorithm/ArrayReverse/SpecExceptionCount}{2}
\pgfkeyssetvalue{/bbe/s2/algorithm/ArrayReverse/SpecFunctionalCount}{2}
\pgfkeyssetvalue{/bbe/s2/algorithm/ArrayReverse/SpecIntermediateCount}{3}
\pgfkeyssetvalue{/bbe/s2/instance/Counter/ConversionTime}{1199.6}
\pgfkeyssetvalue{/bbe/s2/instance/Counter/ConversionOverhead}{0.0531290775988785}
\pgfkeyssetvalue{/bbe/s2/instance/Counter/VerificationTime}{829.0}
\pgfkeyssetvalue{/bbe/s2/instance/Counter/SourceLinesOfCode}{48}
\pgfkeyssetvalue{/bbe/s2/instance/Counter/BytecodeLinesOfCode}{89}
\pgfkeyssetvalue{/bbe/s2/instance/Counter/BoogieLinesOfCode}{183}
\pgfkeyssetvalue{/bbe/s2/instance/Counter/MethodCount}{8}
\pgfkeyssetvalue{/bbe/s2/instance/Counter/SpecRequireCount}{0}
\pgfkeyssetvalue{/bbe/s2/instance/Counter/SpecEnsureCount}{3}
\pgfkeyssetvalue{/bbe/s2/instance/Counter/SpecRaiseCount}{0}
\pgfkeyssetvalue{/bbe/s2/instance/Counter/SpecReturnCount}{4}
\pgfkeyssetvalue{/bbe/s2/instance/Counter/SpecPredicateCount}{3}
\pgfkeyssetvalue{/bbe/s2/instance/Counter/SpecPureCount}{1}
\pgfkeyssetvalue{/bbe/s2/instance/Counter/SpecAssertionCount}{0}
\pgfkeyssetvalue{/bbe/s2/instance/Counter/SpecAssumptionCount}{0}
\pgfkeyssetvalue{/bbe/s2/instance/Counter/SpecInvariantCount}{2}
\pgfkeyssetvalue{/bbe/s2/instance/Counter/SpecExceptionCount}{4}
\pgfkeyssetvalue{/bbe/s2/instance/Counter/SpecFunctionalCount}{3}
\pgfkeyssetvalue{/bbe/s2/instance/Counter/SpecIntermediateCount}{2}
\pgfkeyssetvalue{/bbe/s2/algorithm/GCD/ConversionTime}{1107.8}
\pgfkeyssetvalue{/bbe/s2/algorithm/GCD/ConversionOverhead}{0.0590287195293059}
\pgfkeyssetvalue{/bbe/s2/algorithm/GCD/VerificationTime}{780.8}
\pgfkeyssetvalue{/bbe/s2/algorithm/GCD/SourceLinesOfCode}{51}
\pgfkeyssetvalue{/bbe/s2/algorithm/GCD/BytecodeLinesOfCode}{110}
\pgfkeyssetvalue{/bbe/s2/algorithm/GCD/BoogieLinesOfCode}{203}
\pgfkeyssetvalue{/bbe/s2/algorithm/GCD/MethodCount}{5}
\pgfkeyssetvalue{/bbe/s2/algorithm/GCD/SpecRequireCount}{0}
\pgfkeyssetvalue{/bbe/s2/algorithm/GCD/SpecEnsureCount}{1}
\pgfkeyssetvalue{/bbe/s2/algorithm/GCD/SpecRaiseCount}{1}
\pgfkeyssetvalue{/bbe/s2/algorithm/GCD/SpecReturnCount}{0}
\pgfkeyssetvalue{/bbe/s2/algorithm/GCD/SpecPredicateCount}{2}
\pgfkeyssetvalue{/bbe/s2/algorithm/GCD/SpecPureCount}{3}
\pgfkeyssetvalue{/bbe/s2/algorithm/GCD/SpecAssertionCount}{0}
\pgfkeyssetvalue{/bbe/s2/algorithm/GCD/SpecAssumptionCount}{0}
\pgfkeyssetvalue{/bbe/s2/algorithm/GCD/SpecInvariantCount}{2}
\pgfkeyssetvalue{/bbe/s2/algorithm/GCD/SpecExceptionCount}{1}
\pgfkeyssetvalue{/bbe/s2/algorithm/GCD/SpecFunctionalCount}{1}
\pgfkeyssetvalue{/bbe/s2/algorithm/GCD/SpecIntermediateCount}{2}
\pgfkeyssetvalue{/bbe/s2/algorithm/LinearSearch/ConversionTime}{1048.4}
\pgfkeyssetvalue{/bbe/s2/algorithm/LinearSearch/ConversionOverhead}{0.0589038855575172}
\pgfkeyssetvalue{/bbe/s2/algorithm/LinearSearch/VerificationTime}{798.2}
\pgfkeyssetvalue{/bbe/s2/algorithm/LinearSearch/SourceLinesOfCode}{46}
\pgfkeyssetvalue{/bbe/s2/algorithm/LinearSearch/BytecodeLinesOfCode}{73}
\pgfkeyssetvalue{/bbe/s2/algorithm/LinearSearch/BoogieLinesOfCode}{155}
\pgfkeyssetvalue{/bbe/s2/algorithm/LinearSearch/MethodCount}{6}
\pgfkeyssetvalue{/bbe/s2/algorithm/LinearSearch/SpecRequireCount}{2}
\pgfkeyssetvalue{/bbe/s2/algorithm/LinearSearch/SpecEnsureCount}{1}
\pgfkeyssetvalue{/bbe/s2/algorithm/LinearSearch/SpecRaiseCount}{0}
\pgfkeyssetvalue{/bbe/s2/algorithm/LinearSearch/SpecReturnCount}{1}
\pgfkeyssetvalue{/bbe/s2/algorithm/LinearSearch/SpecPredicateCount}{4}
\pgfkeyssetvalue{/bbe/s2/algorithm/LinearSearch/SpecPureCount}{0}
\pgfkeyssetvalue{/bbe/s2/algorithm/LinearSearch/SpecAssertionCount}{0}
\pgfkeyssetvalue{/bbe/s2/algorithm/LinearSearch/SpecAssumptionCount}{0}
\pgfkeyssetvalue{/bbe/s2/algorithm/LinearSearch/SpecInvariantCount}{1}
\pgfkeyssetvalue{/bbe/s2/algorithm/LinearSearch/SpecExceptionCount}{1}
\pgfkeyssetvalue{/bbe/s2/algorithm/LinearSearch/SpecFunctionalCount}{3}
\pgfkeyssetvalue{/bbe/s2/algorithm/LinearSearch/SpecIntermediateCount}{1}
\pgfkeyssetvalue{/bbe/k18/exceptions/PotentialIndexOutOfBounds/ConversionTime}{1240.2}
\pgfkeyssetvalue{/bbe/k18/exceptions/PotentialIndexOutOfBounds/ConversionOverhead}{0.0764416095568747}
\pgfkeyssetvalue{/bbe/k18/exceptions/PotentialIndexOutOfBounds/VerificationTime}{822.0}
\pgfkeyssetvalue{/bbe/k18/exceptions/PotentialIndexOutOfBounds/SourceLinesOfCode}{45}
\pgfkeyssetvalue{/bbe/k18/exceptions/PotentialIndexOutOfBounds/BytecodeLinesOfCode}{77}
\pgfkeyssetvalue{/bbe/k18/exceptions/PotentialIndexOutOfBounds/BoogieLinesOfCode}{279}
\pgfkeyssetvalue{/bbe/k18/exceptions/PotentialIndexOutOfBounds/MethodCount}{7}
\pgfkeyssetvalue{/bbe/k18/exceptions/PotentialIndexOutOfBounds/SpecRequireCount}{4}
\pgfkeyssetvalue{/bbe/k18/exceptions/PotentialIndexOutOfBounds/SpecEnsureCount}{0}
\pgfkeyssetvalue{/bbe/k18/exceptions/PotentialIndexOutOfBounds/SpecRaiseCount}{1}
\pgfkeyssetvalue{/bbe/k18/exceptions/PotentialIndexOutOfBounds/SpecReturnCount}{3}
\pgfkeyssetvalue{/bbe/k18/exceptions/PotentialIndexOutOfBounds/SpecPredicateCount}{2}
\pgfkeyssetvalue{/bbe/k18/exceptions/PotentialIndexOutOfBounds/SpecPureCount}{0}
\pgfkeyssetvalue{/bbe/k18/exceptions/PotentialIndexOutOfBounds/SpecAssertionCount}{0}
\pgfkeyssetvalue{/bbe/k18/exceptions/PotentialIndexOutOfBounds/SpecAssumptionCount}{0}
\pgfkeyssetvalue{/bbe/k18/exceptions/PotentialIndexOutOfBounds/SpecInvariantCount}{0}
\pgfkeyssetvalue{/bbe/k18/exceptions/PotentialIndexOutOfBounds/SpecExceptionCount}{4}
\pgfkeyssetvalue{/bbe/k18/exceptions/PotentialIndexOutOfBounds/SpecFunctionalCount}{4}
\pgfkeyssetvalue{/bbe/k18/exceptions/PotentialIndexOutOfBounds/SpecIntermediateCount}{0}
\pgfkeyssetvalue{/bbe/k18/exceptions/PotentialNullDereference/ConversionTime}{1155.6}
\pgfkeyssetvalue{/bbe/k18/exceptions/PotentialNullDereference/ConversionOverhead}{0.0922581377746482}
\pgfkeyssetvalue{/bbe/k18/exceptions/PotentialNullDereference/VerificationTime}{887.4}
\pgfkeyssetvalue{/bbe/k18/exceptions/PotentialNullDereference/SourceLinesOfCode}{41}
\pgfkeyssetvalue{/bbe/k18/exceptions/PotentialNullDereference/BytecodeLinesOfCode}{48}
\pgfkeyssetvalue{/bbe/k18/exceptions/PotentialNullDereference/BoogieLinesOfCode}{309}
\pgfkeyssetvalue{/bbe/k18/exceptions/PotentialNullDereference/MethodCount}{6}
\pgfkeyssetvalue{/bbe/k18/exceptions/PotentialNullDereference/SpecRequireCount}{0}
\pgfkeyssetvalue{/bbe/k18/exceptions/PotentialNullDereference/SpecEnsureCount}{0}
\pgfkeyssetvalue{/bbe/k18/exceptions/PotentialNullDereference/SpecRaiseCount}{0}
\pgfkeyssetvalue{/bbe/k18/exceptions/PotentialNullDereference/SpecReturnCount}{6}
\pgfkeyssetvalue{/bbe/k18/exceptions/PotentialNullDereference/SpecPredicateCount}{1}
\pgfkeyssetvalue{/bbe/k18/exceptions/PotentialNullDereference/SpecPureCount}{0}
\pgfkeyssetvalue{/bbe/k18/exceptions/PotentialNullDereference/SpecAssertionCount}{0}
\pgfkeyssetvalue{/bbe/k18/exceptions/PotentialNullDereference/SpecAssumptionCount}{0}
\pgfkeyssetvalue{/bbe/k18/exceptions/PotentialNullDereference/SpecInvariantCount}{0}
\pgfkeyssetvalue{/bbe/k18/exceptions/PotentialNullDereference/SpecExceptionCount}{6}
\pgfkeyssetvalue{/bbe/k18/exceptions/PotentialNullDereference/SpecFunctionalCount}{0}
\pgfkeyssetvalue{/bbe/k18/exceptions/PotentialNullDereference/SpecIntermediateCount}{0}
\pgfkeyssetvalue{/bbe/k18/exceptions/Basic/ConversionTime}{1164.2}
\pgfkeyssetvalue{/bbe/k18/exceptions/Basic/ConversionOverhead}{0.0961910642327384}
\pgfkeyssetvalue{/bbe/k18/exceptions/Basic/VerificationTime}{1091.4}
\pgfkeyssetvalue{/bbe/k18/exceptions/Basic/SourceLinesOfCode}{121}
\pgfkeyssetvalue{/bbe/k18/exceptions/Basic/BytecodeLinesOfCode}{209}
\pgfkeyssetvalue{/bbe/k18/exceptions/Basic/BoogieLinesOfCode}{442}
\pgfkeyssetvalue{/bbe/k18/exceptions/Basic/MethodCount}{22}
\pgfkeyssetvalue{/bbe/k18/exceptions/Basic/SpecRequireCount}{0}
\pgfkeyssetvalue{/bbe/k18/exceptions/Basic/SpecEnsureCount}{0}
\pgfkeyssetvalue{/bbe/k18/exceptions/Basic/SpecRaiseCount}{6}
\pgfkeyssetvalue{/bbe/k18/exceptions/Basic/SpecReturnCount}{6}
\pgfkeyssetvalue{/bbe/k18/exceptions/Basic/SpecPredicateCount}{7}
\pgfkeyssetvalue{/bbe/k18/exceptions/Basic/SpecPureCount}{6}
\pgfkeyssetvalue{/bbe/k18/exceptions/Basic/SpecAssertionCount}{3}
\pgfkeyssetvalue{/bbe/k18/exceptions/Basic/SpecAssumptionCount}{0}
\pgfkeyssetvalue{/bbe/k18/exceptions/Basic/SpecInvariantCount}{0}
\pgfkeyssetvalue{/bbe/k18/exceptions/Basic/SpecExceptionCount}{12}
\pgfkeyssetvalue{/bbe/k18/exceptions/Basic/SpecFunctionalCount}{0}
\pgfkeyssetvalue{/bbe/k18/exceptions/Basic/SpecIntermediateCount}{3}
\pgfkeyssetvalue{/bbe/k18/exceptions/TryFinally/ConversionTime}{1286.0}
\pgfkeyssetvalue{/bbe/k18/exceptions/TryFinally/ConversionOverhead}{0.0790523770709875}
\pgfkeyssetvalue{/bbe/k18/exceptions/TryFinally/VerificationTime}{1011.6}
\pgfkeyssetvalue{/bbe/k18/exceptions/TryFinally/SourceLinesOfCode}{108}
\pgfkeyssetvalue{/bbe/k18/exceptions/TryFinally/BytecodeLinesOfCode}{216}
\pgfkeyssetvalue{/bbe/k18/exceptions/TryFinally/BoogieLinesOfCode}{409}
\pgfkeyssetvalue{/bbe/k18/exceptions/TryFinally/MethodCount}{15}
\pgfkeyssetvalue{/bbe/k18/exceptions/TryFinally/SpecRequireCount}{0}
\pgfkeyssetvalue{/bbe/k18/exceptions/TryFinally/SpecEnsureCount}{4}
\pgfkeyssetvalue{/bbe/k18/exceptions/TryFinally/SpecRaiseCount}{1}
\pgfkeyssetvalue{/bbe/k18/exceptions/TryFinally/SpecReturnCount}{2}
\pgfkeyssetvalue{/bbe/k18/exceptions/TryFinally/SpecPredicateCount}{2}
\pgfkeyssetvalue{/bbe/k18/exceptions/TryFinally/SpecPureCount}{0}
\pgfkeyssetvalue{/bbe/k18/exceptions/TryFinally/SpecAssertionCount}{9}
\pgfkeyssetvalue{/bbe/k18/exceptions/TryFinally/SpecAssumptionCount}{0}
\pgfkeyssetvalue{/bbe/k18/exceptions/TryFinally/SpecInvariantCount}{0}
\pgfkeyssetvalue{/bbe/k18/exceptions/TryFinally/SpecExceptionCount}{3}
\pgfkeyssetvalue{/bbe/k18/exceptions/TryFinally/SpecFunctionalCount}{4}
\pgfkeyssetvalue{/bbe/k18/exceptions/TryFinally/SpecIntermediateCount}{9}
\pgfkeyssetvalue{/bbe/k18/algorithm/ArrayReverse/ConversionTime}{1222.8}
\pgfkeyssetvalue{/bbe/k18/algorithm/ArrayReverse/ConversionOverhead}{0.0718285449582694}
\pgfkeyssetvalue{/bbe/k18/algorithm/ArrayReverse/VerificationTime}{986.4}
\pgfkeyssetvalue{/bbe/k18/algorithm/ArrayReverse/SourceLinesOfCode}{60}
\pgfkeyssetvalue{/bbe/k18/algorithm/ArrayReverse/BytecodeLinesOfCode}{212}
\pgfkeyssetvalue{/bbe/k18/algorithm/ArrayReverse/BoogieLinesOfCode}{226}
\pgfkeyssetvalue{/bbe/k18/algorithm/ArrayReverse/MethodCount}{8}
\pgfkeyssetvalue{/bbe/k18/algorithm/ArrayReverse/SpecRequireCount}{0}
\pgfkeyssetvalue{/bbe/k18/algorithm/ArrayReverse/SpecEnsureCount}{2}
\pgfkeyssetvalue{/bbe/k18/algorithm/ArrayReverse/SpecRaiseCount}{1}
\pgfkeyssetvalue{/bbe/k18/algorithm/ArrayReverse/SpecReturnCount}{1}
\pgfkeyssetvalue{/bbe/k18/algorithm/ArrayReverse/SpecPredicateCount}{4}
\pgfkeyssetvalue{/bbe/k18/algorithm/ArrayReverse/SpecPureCount}{3}
\pgfkeyssetvalue{/bbe/k18/algorithm/ArrayReverse/SpecAssertionCount}{0}
\pgfkeyssetvalue{/bbe/k18/algorithm/ArrayReverse/SpecAssumptionCount}{0}
\pgfkeyssetvalue{/bbe/k18/algorithm/ArrayReverse/SpecInvariantCount}{3}
\pgfkeyssetvalue{/bbe/k18/algorithm/ArrayReverse/SpecExceptionCount}{2}
\pgfkeyssetvalue{/bbe/k18/algorithm/ArrayReverse/SpecFunctionalCount}{2}
\pgfkeyssetvalue{/bbe/k18/algorithm/ArrayReverse/SpecIntermediateCount}{3}
\pgfkeyssetvalue{/bbe/k18/algorithm/Counter/ConversionTime}{1155.6}
\pgfkeyssetvalue{/bbe/k18/algorithm/Counter/ConversionOverhead}{0.0603099488203514}
\pgfkeyssetvalue{/bbe/k18/algorithm/Counter/VerificationTime}{864.2}
\pgfkeyssetvalue{/bbe/k18/algorithm/Counter/SourceLinesOfCode}{46}
\pgfkeyssetvalue{/bbe/k18/algorithm/Counter/BytecodeLinesOfCode}{86}
\pgfkeyssetvalue{/bbe/k18/algorithm/Counter/BoogieLinesOfCode}{177}
\pgfkeyssetvalue{/bbe/k18/algorithm/Counter/MethodCount}{8}
\pgfkeyssetvalue{/bbe/k18/algorithm/Counter/SpecRequireCount}{0}
\pgfkeyssetvalue{/bbe/k18/algorithm/Counter/SpecEnsureCount}{3}
\pgfkeyssetvalue{/bbe/k18/algorithm/Counter/SpecRaiseCount}{0}
\pgfkeyssetvalue{/bbe/k18/algorithm/Counter/SpecReturnCount}{4}
\pgfkeyssetvalue{/bbe/k18/algorithm/Counter/SpecPredicateCount}{3}
\pgfkeyssetvalue{/bbe/k18/algorithm/Counter/SpecPureCount}{1}
\pgfkeyssetvalue{/bbe/k18/algorithm/Counter/SpecAssertionCount}{0}
\pgfkeyssetvalue{/bbe/k18/algorithm/Counter/SpecAssumptionCount}{0}
\pgfkeyssetvalue{/bbe/k18/algorithm/Counter/SpecInvariantCount}{2}
\pgfkeyssetvalue{/bbe/k18/algorithm/Counter/SpecExceptionCount}{4}
\pgfkeyssetvalue{/bbe/k18/algorithm/Counter/SpecFunctionalCount}{3}
\pgfkeyssetvalue{/bbe/k18/algorithm/Counter/SpecIntermediateCount}{2}
\pgfkeyssetvalue{/bbe/k18/algorithm/GCD/ConversionTime}{1131.8}
\pgfkeyssetvalue{/bbe/k18/algorithm/GCD/ConversionOverhead}{0.0615155867582966}
\pgfkeyssetvalue{/bbe/k18/algorithm/GCD/VerificationTime}{801.0}
\pgfkeyssetvalue{/bbe/k18/algorithm/GCD/SourceLinesOfCode}{50}
\pgfkeyssetvalue{/bbe/k18/algorithm/GCD/BytecodeLinesOfCode}{108}
\pgfkeyssetvalue{/bbe/k18/algorithm/GCD/BoogieLinesOfCode}{202}
\pgfkeyssetvalue{/bbe/k18/algorithm/GCD/MethodCount}{5}
\pgfkeyssetvalue{/bbe/k18/algorithm/GCD/SpecRequireCount}{0}
\pgfkeyssetvalue{/bbe/k18/algorithm/GCD/SpecEnsureCount}{1}
\pgfkeyssetvalue{/bbe/k18/algorithm/GCD/SpecRaiseCount}{1}
\pgfkeyssetvalue{/bbe/k18/algorithm/GCD/SpecReturnCount}{0}
\pgfkeyssetvalue{/bbe/k18/algorithm/GCD/SpecPredicateCount}{2}
\pgfkeyssetvalue{/bbe/k18/algorithm/GCD/SpecPureCount}{3}
\pgfkeyssetvalue{/bbe/k18/algorithm/GCD/SpecAssertionCount}{0}
\pgfkeyssetvalue{/bbe/k18/algorithm/GCD/SpecAssumptionCount}{0}
\pgfkeyssetvalue{/bbe/k18/algorithm/GCD/SpecInvariantCount}{2}
\pgfkeyssetvalue{/bbe/k18/algorithm/GCD/SpecExceptionCount}{1}
\pgfkeyssetvalue{/bbe/k18/algorithm/GCD/SpecFunctionalCount}{1}
\pgfkeyssetvalue{/bbe/k18/algorithm/GCD/SpecIntermediateCount}{2}
\pgfkeyssetvalue{/bbe/k18/algorithm/LinearSearch/ConversionTime}{1262.4}
\pgfkeyssetvalue{/bbe/k18/algorithm/LinearSearch/ConversionOverhead}{0.0634439375775659}
\pgfkeyssetvalue{/bbe/k18/algorithm/LinearSearch/VerificationTime}{777.0}
\pgfkeyssetvalue{/bbe/k18/algorithm/LinearSearch/SourceLinesOfCode}{43}
\pgfkeyssetvalue{/bbe/k18/algorithm/LinearSearch/BytecodeLinesOfCode}{88}
\pgfkeyssetvalue{/bbe/k18/algorithm/LinearSearch/BoogieLinesOfCode}{193}
\pgfkeyssetvalue{/bbe/k18/algorithm/LinearSearch/MethodCount}{6}
\pgfkeyssetvalue{/bbe/k18/algorithm/LinearSearch/SpecRequireCount}{2}
\pgfkeyssetvalue{/bbe/k18/algorithm/LinearSearch/SpecEnsureCount}{1}
\pgfkeyssetvalue{/bbe/k18/algorithm/LinearSearch/SpecRaiseCount}{0}
\pgfkeyssetvalue{/bbe/k18/algorithm/LinearSearch/SpecReturnCount}{1}
\pgfkeyssetvalue{/bbe/k18/algorithm/LinearSearch/SpecPredicateCount}{4}
\pgfkeyssetvalue{/bbe/k18/algorithm/LinearSearch/SpecPureCount}{0}
\pgfkeyssetvalue{/bbe/k18/algorithm/LinearSearch/SpecAssertionCount}{0}
\pgfkeyssetvalue{/bbe/k18/algorithm/LinearSearch/SpecAssumptionCount}{0}
\pgfkeyssetvalue{/bbe/k18/algorithm/LinearSearch/SpecInvariantCount}{1}
\pgfkeyssetvalue{/bbe/k18/algorithm/LinearSearch/SpecExceptionCount}{1}
\pgfkeyssetvalue{/bbe/k18/algorithm/LinearSearch/SpecFunctionalCount}{3}
\pgfkeyssetvalue{/bbe/k18/algorithm/LinearSearch/SpecIntermediateCount}{1}
\pgfkeyssetvalue{/bbe/average/ConversionTime}{1290.945945945946}
\pgfkeyssetvalue{/bbe/average/ConversionOverhead}{0.10585239861485912}
\pgfkeyssetvalue{/bbe/average/VerificationTime}{1188.497297297297}
\pgfkeyssetvalue{/bbe/average/SourceLinesOfCode}{211.0810810810811}
\pgfkeyssetvalue{/bbe/average/BytecodeLinesOfCode}{541.4594594594595}
\pgfkeyssetvalue{/bbe/average/BoogieLinesOfCode}{578.3243243243244}
\pgfkeyssetvalue{/bbe/average/SpecAssertionCount}{2.324324324324324}
\pgfkeyssetvalue{/bbe/average/SpecAssumptionCount}{0.0}
\pgfkeyssetvalue{/bbe/average/SpecInvariantCount}{1.2162162162162162}
\pgfkeyssetvalue{/bbe/average/MethodCount}{28.91891891891892}
\pgfkeyssetvalue{/bbe/average/SpecIntermediateCount}{3.5405405405405403}
\pgfkeyssetvalue{/bbe/average/SpecPredicateCount}{4.216216216216216}
\pgfkeyssetvalue{/bbe/average/SpecFunctionalCount}{2.4054054054054053}
\pgfkeyssetvalue{/bbe/average/SpecExceptionCount}{5.378378378378378}
\pgfkeyssetvalue{/bbe/total/ConversionTime}{47765.00000000001}
\pgfkeyssetvalue{/bbe/total/ConversionOverhead}{3.9165387487497876}
\pgfkeyssetvalue{/bbe/total/VerificationTime}{43974.399999999994}
\pgfkeyssetvalue{/bbe/total/SourceLinesOfCode}{7810}
\pgfkeyssetvalue{/bbe/total/BytecodeLinesOfCode}{20034}
\pgfkeyssetvalue{/bbe/total/BoogieLinesOfCode}{21398}
\pgfkeyssetvalue{/bbe/total/MethodCount}{1070}
\pgfkeyssetvalue{/bbe/total/SpecRequireCount}{30}
\pgfkeyssetvalue{/bbe/total/SpecEnsureCount}{59}
\pgfkeyssetvalue{/bbe/total/SpecRaiseCount}{80}
\pgfkeyssetvalue{/bbe/total/SpecReturnCount}{119}
\pgfkeyssetvalue{/bbe/total/SpecPredicateCount}{156}
\pgfkeyssetvalue{/bbe/total/SpecPureCount}{79}
\pgfkeyssetvalue{/bbe/total/SpecAssertionCount}{86}
\pgfkeyssetvalue{/bbe/total/SpecAssumptionCount}{0}
\pgfkeyssetvalue{/bbe/total/SpecInvariantCount}{45}
\pgfkeyssetvalue{/bbe/total/SpecFunctionalCount}{89}
\pgfkeyssetvalue{/bbe/total/SpecExceptionCount}{199}
\pgfkeyssetvalue{/bbe/total/SpecIntermediateCount}{131}

\ifblind
\author{Anonymous authors}
\else
\author{Marco Paganoni\inst{1} \and
Carlo A. Furia\inst{1}\orcidID{0000-0003-1040-3201}}

\authorrunning{M. Paganoni and C. A. Furia}

\institute{Software Institute, USI Università della Svizzera italiana, Lugano, Switzerland\\
\email{marco.paganoni@usi.ch} $\qquad$ \url{bugcounting.net}}
\fi

\maketitle

\begin{abstract}
  A program's exceptional behavior can substantially complicate its control flow,
  and hence accurately reasoning about the program's correctness.
  On the other hand, formally verifying realistic programs
  is likely to involve exceptions---a ubiquitous feature
  in modern programming languages.
  
  In this paper, we present a novel approach to verify
  the exceptional behavior of Java programs,
  which extends our previous work on \byteback.
  \byteback works on a program's bytecode,
  while providing means to specify the intended behavior
  at the source-code level;
  this approach sets \byteback apart from
  most state-of-the-art verifiers that target source code.
  To explicitly model a program's exceptional behavior
  in a way that is amenable to formal reasoning,
  we introduce Vimp: a high-level bytecode representation
  that extends the Soot framework's Grimp with verification-oriented features,
  thus serving as an intermediate layer between bytecode and the
  Boogie intermediate verification language.
  Working on bytecode through this intermediate layer
  brings flexibility and adaptability
  to new language versions and variants:
  as our experiments demonstrate, \byteback can verify
  programs involving exceptional behavior in all versions of Java,
  as well as in Scala and Kotlin (two other popular JVM languages).
\end{abstract}

\section{Introduction}
\label{sec:introduction}

Nearly every modern programming language
supports exceptions
as a mechanism to signal and handle
unusual runtime conditions
(so-called \emph{exceptional behavior})
separately from the main
control flow
(the program's \emph{normal} behavior).
Exceptions are usually preferable 
to lower-level ad hoc solutions (such as error codes and defensive programming), 
because deploying them does not pollute the source code's structured control flow.
However, by introducing extra, often implicit execution paths, 
exceptions may also complicate reasoning about all possible program behavior---and thus, ultimately, about program correctness.

In this paper, 
we introduce a novel approach to perform deductive verification
of Java programs involving exceptional behavior.
Exceptions were baked into the Java programming language
since its inception,
where
they remain widely used~\cite{AnalysisOfPatterns-MSR16,UnveilingException-SANER19,MF-MSR21};
nevertheless,
as the example of \autoref{sec:examples} demonstrates,
they can be somewhat of a challenge to reason about formally.
To model together normal and exceptional control flow paths,
and to seamlessly support any exception-related language features, 
our verification approach
crucially 
targets a program's \emph{bytecode} intermediate representation---instead of the source code analyzed by state-of-the-art verifiers
such as KeY~\cite{KeY} and OpenJML~\cite{OpenJML}.
We introduced the idea of performing formal verification
at the level of bytecode in previous work~\cite{PF-FM23-ByteBack}.
In this paper,
we build on those results
and implement support for exceptions in the \byteback deductive verifier.

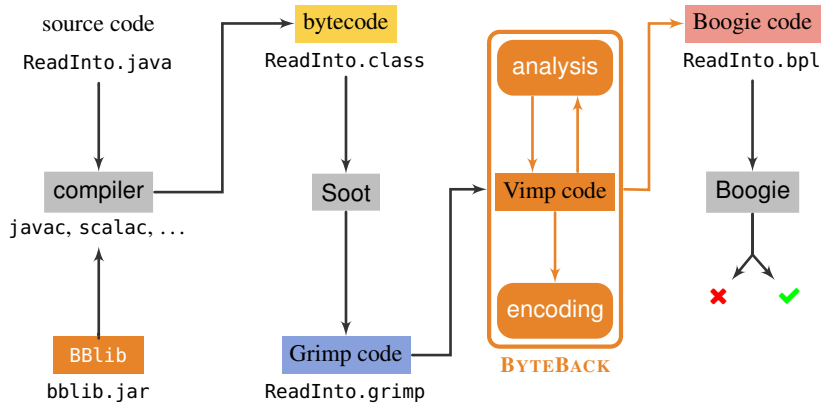
\begin{figure}[!bt]
\begin{adjustwidth}{-30mm}{-30mm}
\centering
\begin{tikzpicture}[
  bbbox/.style={rectangle,very thick,
    rounded corners=2mm,font=\footnotesize\sffamily,
    minimum width=15mm,minimum height=7mm,
    draw=bbcol,fill=bbcol,text=white,
    label={[below=15pt]#1}},
  toolbox/.style 2 args={rectangle,very thick,
    font=\footnotesize\sffamily,
    draw=black!25,fill=black!25,text=black,
    label={[below=15pt,#2]#1}},  databox/.style 2 args={font=\footnotesize,
    label={[below=15pt,#2]#1}},  align=center,
  node distance=12mm and 20mm,
  ]

  \matrix[row sep=12mm,column sep=8mm] {
\node[databox={\texttt{ReadInto.java}}{name=source-lab}] (source) {source code};
    &
    \node[databox={\texttt{ReadInto.class}}{name=bytecode-lab},fill=bytecodecol] (bytecode) {bytecode};
    &
    \node (bb-top) {};
    & 
    \node[databox={\texttt{ReadInto.bpl}}{name=boogie-code-lab},fill=boogiecol] (boogie-code) {Boogie code};
\\
    \node[toolbox={\texttt{javac},\ \texttt{scalac},\ \ldots}{name=compiler-lab}] (compiler) {compiler};
    &
    \node[toolbox] (soot) {Soot};
    &
    \node[databox={}{name=vimp-code-lab},fill=bbcol] (bb-vimp) {Vimp code};
    &
    \node[toolbox] (boogie) {Boogie};
    \\
    \node[databox={\texttt{bblib.jar}}{name=bblib-lab},fill=bbcol,text=white,
          minimum width=12mm,minimum height=5mm] (bblib) {\bblib};
    &
    \node[databox={\texttt{ReadInto.grimp}}{name=grimp-code-lab},fill=jimplecol] (grimp) {Grimp code};
    &
    \node[] (bb-bottom) {};
    \\
  };

  \node[bbbox,below=3pt of bb-top] (analysis) {analysis};
  \node[bbbox,above=3pt of bb-bottom] (encoding) {encoding};
  \node[below=10mm of boogie] (outcome) {};
  \node[right=3pt of outcome] (ok) {\color{green}{\faCheck}};
  \node[left=3pt of outcome] (fail) {\color{red}{\faClose}};

  \node [fit=(analysis)(encoding),draw=bbcol,
         ultra thick,rounded corners,label={[bbcol]south:\textbf{\byteback}}] (byteback) {};

   \coordinate (mid-left) at ($(soot)!0.5!(byteback)$);
   \coordinate (bot-mid-left) at ($(mid-left)!(grimp)!(mid-left |- grimp)$);
   \coordinate (an-left) at ($(analysis.south)+(-3mm, 0)$);
   \coordinate (an-right) at ($(analysis.south)+(+3mm, 0)$);
         
  \begin{scope}[color=black!80,line width=1pt,round cap-latex',every node/.style={font=\footnotesize}]
    \draw (source-lab) -- (compiler);
    \draw (compiler.east) -- ($(compiler)!0.5!(soot)$) |- (bytecode.west);
    \draw (bytecode-lab) -- (soot);
    \draw (soot) -- (grimp);
    \draw (grimp.east) -- (bot-mid-left) |- (byteback);
    \draw[color=bbcol] (bb-vimp) -- (encoding);
    \draw[color=bbcol] (an-left) -- (an-left |- bb-vimp.north);
    \draw[color=bbcol] (an-right |- bb-vimp.north) -- (an-right);
    \draw[color=bbcol] (byteback.east) -- ($(byteback.east)!0.2!(boogie)$) |- (boogie-code.west);
    \draw (boogie-code-lab) -- (boogie);
    \draw (boogie) -- ($(boogie)!0.6!(outcome)$) -- (ok);
    \draw (boogie) -- ($(boogie)!0.6!(outcome)$) -- (fail);

    \draw (bblib.north) -| (compiler-lab);
\end{scope}
  
\end{tikzpicture}
\end{adjustwidth}
\caption{An overview of \byteback's verification workflow.}
\label{fig:workflow}
\end{figure}

The key idea of our \byteback approach
(pictured in \autoref{fig:workflow})
is using JVM bytecode solely
as a convenient intermediate representation;
users of \byteback 
still annotate program source code
in a very similar way as if they were working with a source-level verifier.
To this end,
we extend the
specification library introduced with \byteback
(called \bblib)
with features 
to specify exceptional behavior
(for example, conditions under which a method terminates normally or exceptionally)
using custom Java expressions;
thus,
such specifications
remain available in bytecode
after compiling an annotated program
using a standard Java compiler.
\byteback analyzes the bytecode
and encodes the program's semantics,
its specification, and other information necessary for verification
into
Boogie~\cite{Boogie}---a widely-used intermediate language for verification;
then, verifying the Boogie translation is equivalent
to verifying the original Java program against its specification.

As we demonstrate with experiments in \autoref{sec:experiments},
performing verification on bytecode offers several advantages:
\begin{enumerate*}
\item \emph{Robustness} to source-language changes:
  while Java evolves rapidly, frequently introducing new features
  (also for exceptional behavior),
  bytecode is generally stable;
  thus, our verification technique continues to work with the latest Java versions.
\item \emph{Multi-language} support:
  \byteback and its \bblib specification library
  are designed so that they can be applied, in principle, to specify
  programs in any language that is bytecode-compatible;
  while the bulk of our examples are in Java,
  we will demonstrate verifying exceptional behavior in Scala and Kotlin---two modern languages for the JVM.
\item \emph{Flexibility} of modeling:
  since exceptional behavior becomes explicit in bytecode,
  the \byteback approach extensively and seamlessly deals with
  any intricate exceptional behavior (such as implicit or suppressed exceptions).
\end{enumerate*}

\nicepar{Contributions and positioning.}
In summary, the paper makes the following contributions:
\begin{enumerate*}
\item Specification features to specify exceptional behavior of JVM languages.
\item A verification technique that encodes bytecode exceptional behavior into Boogie.
\item An implementation of the specification features and the verification technique
  that extend the \bblib library and \byteback verifier.
\item Vimp: a high-level bytecode format
  suitable to reason about functional correctness, built
  on top of Soot's Grimp format~\cite{Soot,Soot2}.
\item An experimental evaluation with \n.{count}! programs involving exceptional behavior in Java, Scala, and Kotlin.
\item For reproducibility, \byteback and all experimental artifacts are available in a replication package~\cite{Byteback-REPP}.
\end{enumerate*}
While we build upon \bblib and \byteback,
introduced in previous work of ours~\cite{PF-FM23-ByteBack},
this paper's contributions substantially extend
them with support for exceptional behavior,
as well as other Java related features (see \autoref{sec:methodology:spec-verify-except}).
For simplicity, henceforth ``\bblib'' and ``\byteback''
denote their current versions, equipped with the novel contributions
described in the rest of the paper.

\section{Motivating Example}
\label{sec:examples}

Exceptions can significantly complicate
the control flow of even seemingly simple code;
consequently, correctly reasoning about exceptional behavior
can be challenging even for a language like Java---whose exception-handling features have not changed significantly
since the language's origins.

\begin{figure}[!bt]
  \centering
  \begin{subfigure}[t]{0.53\linewidth}
\begin{lstlisting}[language=BBlib]
@Require(r == null || !r.closed)
@Raise(NullPointerException, r == null)
@Return(a != null && r != null)
@Ensure(r == null || r.closed)
static void into(final Resource r, final int[] a) {
  try (r) {
    int i = 0;
    while (true) { invariant(0 <= i <= a.length);
                   invariant(r == null || !r.closed);
      a[i] = r.read(); ++i;
    }
  } catch (IndexOutOfBoundsException
           | NoSuchElementException e) { return; }
}
\end{lstlisting}
    \caption{Method \J{into} copies \J{r}'s content into array \J{a}. It is annotated with normal and exceptional pre- and postconditions using a simplified \bblib syntax.}
    \label{code:into}
  \end{subfigure}
\hfill
  \begin{subfigure}[t]{0.44\linewidth}
\begin{lstlisting}[language=BBlib,numbers=none]
class Resource implements AutoCloseable {

   boolean closed;
   boolean hasNext;

   @Raise(IllegalStateException, closed)
   @Raise(NoSuchElementException, !hasNext)
   @Return(!closed && hasNext)
   int read()
   { /* ... */ }

   // ...

}
\end{lstlisting}
    \caption{An outline of class \J{Resouce}'s interface with Boolean attributes \J{closed} and \J{hasNext}, and method \J{read}.}
    \label{code:resource}
  \end{subfigure}
  \caption{Annotated Java method \J{into} and class \J{Resource}, which demonstrate some pitfalls of specifying and reasoning about exceptional behavior.}
  \label{fig:motivating-examples}
\end{figure}

To demonstrate, consider \autoref{code:into}'s method \J{into},
which inputs a reference \J{r} to a \J{Resource} object
and an integer array \J{a},
and copies values from \J{r} into \J{a}---until either \J{a} is filled up or there are no more values in \J{r} to read.
\autoref{code:resource} shows the key features of class \J{Resource}---which implements Java's \J{AutoCloseable} interface,
and hence can be used similarly to most standard I/O classes.
Method \J{into}'s implementation uses a \J{try}-with-resources
block to ensure that \J{r} is closed whenever the method terminates---normally or exceptionally.
The \J{while} loop terminates as soon as any of the following
conditions holds:
\begin{enumerate*}
\item \label{case:npe} \J{r} is \J{null};
\item \label{case:outofbounds} \J{k} reaches \J{a.length} (array \J{a} is full);
\item \label{case:noelement} \J{r.read()} throws a \J{NoSuchElementException} (\J{r} has no more elements).
\end{enumerate*}
Method \J{into} returns with an (propagated) exception 
only in case \ref{case:npe};
case \ref{case:outofbounds}'s \J{IndexOutOfBounds}
and case \ref{case:noelement}'s \J{NoSuchElement} exceptions
are caught by the \J{catch} block that \emph{suppresses} them with a \J{return}---so that \J{into} can return normally.

\autoref{code:into}'s annotations,
which use a simplified syntax for \bblib ---\byteback's specification library---specify part of \J{into}'s expected behavior.
Precondition \BBl{@Require} expresses
the constraint that object \J{r} must be open (or \J{null})
for \J{into} to work as intended.
Annotation \BBl{@Raise} declares that
\J{into} terminates with a \J{NullPointer} exception
if \J{r} is \J{null};
conversely, \BBl{@Return} says that \J{into} returns \emph{normally}
if \J{a} and \J{r} are not \J{null}.
Finally, postcondition \BBl{@Ensure}
says that, if it's not \J{null}, \J{r} will be closed when \J{into} terminates---regardless of whether it does so normally or exceptionally. 
The combination of language features and numerous forking control-flow paths
complicate reasoning about---and even specifying---\J{into}'s behavior.
While exception handling has been part of Java since version~1,
try-with-resources and multi-catch (both used in \autoref{code:into})
were introduced in Java~7;
thus, even a state-of-the-art verifier like KeY~\cite{KeY} lacks support for them.
OpenJML~\cite{OpenJML} can reason about all exception features up to Java~7;
however, the try-with-resources using an existing \J{final} variable
became available only in Java~9.\urlcite{https://jcp.org/en/jsr/detail?id=334}
Furthermore,
OpenJML implicitly checks that verified code
does not throw any implicit exceptions (such as \J{NullPointer} or \J{IndexOutOfBound} exceptions)---thus disallowing code such as \autoref{code:into}'s,
where implicitly throwing exceptions is part of the expected behavior.
These observations are best thought of as 
design choices---rather than limitations---of these powerful Java verification tools:
after all, features such as multi-catch
are syntactic sugar that makes programs
more concise but does not affect expressiveness;
and propagating uncaught implicit exceptions
can be considered an
anti-pattern\urlcite{https://docs.oracle.com/javase/tutorial/essential/exceptions/runtime.html}~\cite{WeimerN04,MF-ICSME22-Wit}. 
However, they also speak volumes to the difficulty of
fully supporting all cases of exceptional behavior in a feature-laden
language like Java.

As we detail in the rest of the paper,
\byteback's approach offers advantages in such scenarios.
Crucially, the implicit control flow of exception-handling code
becomes explicit when compiled to bytecode,
which eases analyzing it consistently
and disentangling the various specification elements.
For instance,
the \J{while} loop's several exceptional exit points
become apparent in \J{into}'s bytecode translation,
and \byteback can check that the declared invariant holds in all of them,
and that postcondition \BBl{@Ensure} holds in all matching method return points.
Furthermore, bytecode is more stable than Java---thus, a verifier like \byteback is more robust to source language evolution.
Thanks to these capabilities, \byteback can verify the behavior of \autoref{fig:motivating-examples}'s example.

\section{Specifying and Verifying Exceptional Behavior}
\label{sec:methodology:spec-verify-except}

This section describes the new features of \byteback
to specify and verify exceptional behavior.
\autoref{fig:workflow}
shows \byteback's workflow,
which we revisited to support these new features.
Users of \byteback ---just like with every deductive verifier---have to annotate the source code to be verified
with a specification and other annotations.
To this end, \byteback offers \bblib:
an annotation library
that is usable with any language that is bytecode compatible.
\autoref{subsec:byteback-spec} describes
the new \bblib features to specify exceptional behavior.
Then, users compile the
annotated source code 
with the language's bytecode compiler.
\byteback relies on the Soot static analysis framework
to analyze bytecode;
precisely, Soot offers Grimp:
a higher-level alternative bytecode representation.
\byteback processes Grimp and translates it to Vimp:
a verification-oriented extension of Grimp
that we introduced in the latest \byteback version
and is described in \autoref{subsec:vimp}.
As we discuss in \autoref{subsec:exceptional-control-flow},
\byteback transforms Grimp to Vimp
in steps, each taking care of a different aspect
(expressions, types, control flow, and so on).
Once the transformation is complete,
\byteback encodes the Vimp program into
the Boogie intermediate verification language~\cite{Boogie};
thanks to Vimp's custom design, the Boogie encoding is mostly straightforward.
Finally, the Boogie tool verifies the generated Boogie program,
and reports success or any verification failures
(which can be manually
traced back to source-code specifications that could not be verified).

\subsection{Specifying Exceptional Behavior}
\label{subsec:byteback-spec}

Users of \byteback add behavioral specifications to
a program's source using \bblib:
\byteback's
standalone Java library, 
offering annotation tags and static methods
suitable to specify functional behavior.
Since \bblib uses only basic language constructs,
it is compatible with most JVM languages
(as we demonstrate in \autoref{sec:experiments});
and all the information added as \bblib annotations
is preserved at the bytecode level.
This section first summarizes the core characteristics of \bblib
(to make the paper self contained),
and then describes the features
we introduced to specify exceptional behavior.

\subsubsection{Specification Expressions}
Expressions used in \byteback specifications
must be \emph{aggregable},
that is
pure
(they can be evaluated without side effects)
and branchless
(they can be evaluated without branching instructions).
These are common requirements
to ensure that specification expressions
are well-formed~\cite{RudichDM08}---hence, 
equivalently expressible as purely \emph{logic expressions}.
Correspondingly, \bblib forbids impure expressions,
and offers aggregable replacements
for the many Java operators that introduce branches
in the bytecode,
such as the standard Boolean operators (\J{&&}, \J{||}, \ldots)
and comparison operators (\J{==}, \J{<}, \ldots).
\autoref{tab:bblib-ops} shows several of \bblib's aggregable operators,
including some that have no immediately equivalent Java expression
(such as the quantifiers).
Using only \bblib's operators in a specification ensures
that it remains in a form that \byteback can process
after the source program has been compiled to bytecode~\cite{PF-FM23-ByteBack}.
Thus,
\bblib operators map to Vimp logic operators (\autoref{subsec:vimp}),
which directly translate to Boogie operators with matching semantics (\autoref{sec:trasformation-order}). 

\begin{table}[!htb]
  \centering
  \setlength{\tabcolsep}{6pt}
  \footnotesize
  \begin{tabular}{lcc}
    \toprule
    & \textsc{in Java/logic} & \textsc{in \bblib} \\
    \midrule
    \multirow{2}{*}{comparison}
    & \J{x < y}, \J{x <= y}, \J{x == y} & \BBl{lt(x, y)}, \BBl{lte(x, y)}, \BBl{eq(x, y)}
    \\
    & \J{x != y}, \J{x >= y}, \J{x > y} & \BBl{neq(x, y)}, \BBl{gte(x, y)}, \BBl{gt(x, y)}
    \\[1pt]
    conditionals & \J{c ? t : e} & \BBl{conditional(c, t, e)}
    \\
    \cmidrule(lr){2-3}
    propositional
    & \J{!a}, \J{a && b}, \J{a \|$\!$\| b}, \J{a$\ \Longrightarrow$ b}
    & \BBl{not(a)}, \BBl{a & b}, \BBl{a \| b}, \BBl{implies(a, b)}
    \\[1pt]
    \multirow{2}{*}{quantifiers} & \BBl{$\forall$x: T :: P(x)}& \BBl{T x = Binding.T(); forall(x, P(x))}
    \\
    & \BBl{$\exists$x: T :: P(x)}& \BBl{T x = Binding.T(); exists(x, P(x))}
    \\
    \bottomrule
  \end{tabular}
  \caption{\bblib's aggregable operators, used instead of Java's impure or branching operators in specification expressions.}
  \label{tab:bblib-ops}
\end{table}

\subsubsection{Method Specifications}
To specify the input/output behavior of methods,
\bblib offers annotations \BBl{@Require} and \BBl{@Ensure}
to express a method's pre- and postconditions.
For example, \BBl{@Require($p$) @Ensure($q$) t m(args)}
specifies that $p$ and $q$ are method \J{m}'s pre- and postcondition;
both $p$ and $q$ denote names of \emph{predicate} methods,
which are methods marked with annotation \BBl{@Predicate}.
As part of the verification process,
\byteback checks that every such predicate method $p$ is well-formed:
\begin{enumerate*}
\item $p$ returns a \J{boolean};
\item $p$'s signature is the same as \J{m}'s or,
  if $p$ is used in \J{m}'s postcondition and \J{m}'s return type is not \J{void},
  it also includes an additional argument that denotes \J{m}'s returned value;
\item $p$'s body returns a single aggregable expression;
\item if $p$'s body calls other methods, they must also be aggregable.
\end{enumerate*}
For example, the postcondition $q$ of a method \BBl{int fun(int x)}
that always returns a value greater than the input \J{x}
is expressible in \bblib as
\BBl{@Predicate boolean $\ q$(int x, int result) \{return gt(result, x);\}}.
Postcondition predicates may also
refer to a method's pre-state
by means of \BBl{old($a$)} expressions, 
which evaluates to $a$'s value in the method's pre-state.

For simplicity of presentation,
we henceforth abuse the notation and use an identifier
to denote a predicate method's \emph{name},
the \emph{declaration} of the predicate method,
and the \emph{expression} that the predicate method returns.
Consider, for instance, \autoref{code:into}'s precondition;
in \bblib syntax,
it can be declared as \BBl{@Require("null_or_open")},
where \J{"null_or_open"} is the name of a method \J{null_or_open},
whose body returns expression \BBl{eq(r, null)$\,$\|$\,$not(r.closed)},
which is \J{into}'s actual precondition.

\subsubsection{Exceptional Postconditions}

Predicate methods
specified with \BBl{@Ensure}
are evaluated on a method's post-state
regardless of whether the method terminated normally or with an exception;
for example, predicate \J{x_eq_y} in \autoref{code:exceptional-postcondition}
says that attribute \J{x} always equals argument \J{y} when method \J{m} terminates.
To specify exceptional behavior,
a method's postcondition predicate 
may include an additional argument \J{e} of type \J{Throwable},
which denotes the thrown exception if the method terminated with an exception
or satisfies \bblib's predicate \J{isVoid(e)} if the method terminated normally.
Predicate \J{x_pos} 
in \autoref{code:exceptional-postcondition}
is an example of exceptional behavior,
as it says that \J{m} throws an exception of type \J{PosXExc}
when attribute \J{x} is greater than zero upon termination;
conversely, predicate \J{y_neg} specifies that \J{m} terminates normally
when argument \J{y} is negative or zero.

\begin{figure}[!bt]
\begin{lstlisting}[language=BBlib,basicstyle=\ttfamily\small]
class C {

  int x = 0;

  @Ensure("x_eq_y")  // x == y when m terminates normally or exceptionally
  @Ensure("x_pos")   // if x > 0 then m throws an exception
  @Ensure("y_neg")   // if y <= 0 then m terminates normally
  void m(int y) { x = y; if (x > 0) throw new PosXExc(); }

  @Predicate public boolean y_neg(int y, Throwable e)
  { return implies(lte(y, 0), isVoid(e)); }

  @Predicate public boolean x_pos(int y, Throwable e)
  { return implies(gt(x, 0), e instanceof PosXExc); }

  @Predicate public boolean x_eq_y(int y)
  { return eq(x, y); }

}
\end{lstlisting}
  \caption{Examples of exceptional postconditions in \bblib.}
  \label{code:exceptional-postcondition}
\end{figure}

\begin{table}
  \centering
  \begin{tabular}{l l}
    \toprule
    \multicolumn{1}{c}{\textsc{shorthand}}         & \multicolumn{1}{c}{\textsc{equivalent postcondition}} \\
    \midrule 
    \BBl{@Raise(exception = E.class, when =\ $p$)} & \BBl{@Ensure(implies(old($p$),\ $e$ instanceof E))}   \\
    \BBl{@Return(when =\ $p$)}                     & \BBl{@Ensure(implies(old($p$), isVoid($e$)))}         \\
    \BBl{@Return}                                  & \BBl{@Return(when = true)}                            \\
    \bottomrule
  \end{tabular}
  \caption{\bblib's \BBl{@Raise} and \BBl{@Return} and annotation shorthands.}
  \label{tab:annotation-shorthands}
\end{table}

\subsubsection{Shorthands}
For convenience, \bblib offers annotation shorthands \BBl{@Raise} and \BBl{@Return}
to specify when a method terminates exceptionally or normally.
\autoref{tab:annotation-shorthands} shows
the semantics of these shorthands by translating them into equivalent postconditions.
The \J{when} argument refers to a method's pre-state through the \BBl{old} expression,
since it is common to relate a method's exceptional behavior to its inputs.
Thus, \autoref{code:exceptional-postcondition}'s postcondition \J{y_neg}
is equivalent to \BBl{@Return(lte(y, 0))}, since \J{m} does not change \J{y}'s value;
conversely, \BBl{@Raise(PosXExc.class, gt(x, 0))}
is \emph{not} equivalent to \J{x_pos} because \J{m} sets \J{x} to \J{y}'s value.

\subsubsection{Intermediate Specification}
\bblib also supports the usual intra-method specification elements:
assertions, assumptions, and loop invariants.
Given an aggregable expression $e$,
\BBl{assertion($e$)}
specifies that $e$ must hold whenever execution reaches it;
\BBl{assumption($e$)}
restricts verification from this point on to only executions where $e$ holds;
and \BBl{invariant($e$)}
declares that $e$ is an invariant of the loop within whose body it is declared.
As we have seen in \autoref{code:into}'s running example,
loop invariants hold, in particular, at all exit points of a loop---including exceptional ones.

\subsection{The Vimp Intermediate Representation}
\label{subsec:vimp}

In our previous work~\cite{PF-FM23-ByteBack},
\byteback works directly on Grimp---a high-level bytecode
representation provided by the Soot static analysis framework~\cite{Soot,Soot2}.
Compared to raw bytecode,
Grimp conveniently retains information such as types and expressions,
which eases \byteback's encoding of the program under verification into Boogie~\cite{Boogie}.
However, Grimp remains a form of bytecode,
and hence it represents well executable instructions,
but lacks support
for encoding logic expressions and specification constructs.
These limitations become especially inconvenient
when reasoning about exceptional behavior,
which often involves logic conditions
that depend on the types and values of exceptional objects.
Rather than reconstructing this information during the translation from
Grimp to Boogie,
we found it more effective to extend Grimp into Vimp, 
which fully supports logic and specification expressions.

Our bespoke Vimp bytecode representation 
can encode all the information relevant for verification.
This brings several advantages:
\begin{enumerate*}
\item it decouples the input program's static analysis from the generation of Boogie code, achieving more flexibility at either ends of the toolchain;
\item it makes the generation of Boogie code straightforward (mostly one-to-one);
\item \byteback's transformation from Grimp to Vimp becomes naturally \emph{modular}:
  it composes several simpler transformations, each taking care of a different aspect
  and incorporating a different kind of information.
\end{enumerate*} 
The rest of this section presents Vimp's key features,
and how they are used by \byteback's Grimp-to-Vimp transformation \tr.\footnote{In the following, we occasionally take some liberties with Grimp and Vimp code,
  using a readable syntax
  that mixes bytecode instruction and Java statement syntax;
  for
  example, \Gr{m()} represents an invocation of method \Gr{m}
  that corresponds to a suitable variant of bytecode's \Gr{invoke}.
}
As detailed in \autoref{sec:trasformation-order},
\tr composes the following feature-specific transformations: 
\tr[exc] makes the exceptional control flow explicit;
\tr[agg] aggregates Grimp expressions into compound Vimp expressions;
\tr[inst] translates Grimp instructions
(by applying
transformation \tr[stm] to statements,
transformation \tr[exp] to expressions within statements,
and \tr[types] to expression types);
\tr[loop] handles loop invariants.

\subsubsection{Expression Aggregation}
Transformation \tr[agg]
\emph{aggregates} specification expressions (see~\autoref{subsec:byteback-spec}),
so that each corresponds to a single Vimp pure and branchless expression.
In a nutshell, $\tr[agg](s)$
takes a piece $s$ of aggregable Grimp code,
converts it into static-single assignment form,
and then recursively replaces each variable's single usage
with its unique definition.
For example, consider \autoref{code:into}'s loop invariant:
it corresponds to \Gr{a := lte(0, i); b := lte(i, a.length); c := a \& b} in Grimp,
and becomes \Vi{c := lte(0, i) \& lte(i, a.length)} in Vimp.

\subsubsection{Expected Types}
Transformation $\tr[type]$
reconstructs the \emph{expected type} of expressions
when translating them to Vimp.
An expression $e$'s expected type
depends on the context where $e$ is used;
in general, it differs from $e$'s type in Grimp,
since Soot's type inference
may not distinguish between Boolean and integer expressions---which both use the \J{int} bytecode type.

\subsubsection{Boolean Expressions}
Another consequence of bytecode's lack of a proper \J{boolean} representation
is that Grimp uses integer operators also as Boolean operators
(for example the unary minus \Gr{-} for ``not'').
In contrast,
Vimp supports the usual Boolean operators
\Vi{!}, \Vi{&&}, \Vi{\|\|}, \Vi{==>},
and constants \Vi{true} and \Vi{false}.
Transformation \tr[exp] uses them to translate 
Vimp expressions $e$ whose expected type $\tr[type](e)$ is \Vi{boolean};
this includes specification expressions (which use \bblib's replacement operators),
but also regular Boolean expressions in the executable code.
For example,
$\tr[exp](\Gr{-a}) = \Vi{!}\!\!\!\tr[exp](a)$,
$\tr[exp](k) = \Vi{true}$ for every constant $k \geq 1$,
and $\tr[exp](h) = \Vi{false}$ for every constant $h < 1$.

Transformation $\tr[exp](e)$
also identifies quantified expressions---expressed using a combination of \bblib's \BBl{Contract} and \BBl{Binding}---after aggregating them,
and renders them using Vimp's quantifier syntax:
\begin{small}
\begin{align*}
  \tr[exp](\Gr{Contract.forall(Binding.T(), }e\Gr{)})
  &=\ \Vi{forall}\;\tr[type](\Gr{Binding.T()})\:\Vi{v :: }\tr[exp](e)
    \\
  \tr[exp](\Gr{Contract.exists(Binding.T(), }e\Gr{)})
  &=\ \Vi{exists}\;\tr[type](\Gr{Binding.T()})\:\Vi{v :: }\tr[exp](e)
\end{align*}
\end{small}

\subsubsection{Assertion Instructions}
\label{subsubsec:assertion-instructions}

Vimp includes instructions \Vi{assert}, \Vi{assume}, and \Vi{invariant},
which transformation \tr[stm]
introduces for each corresponding instance of
\bblib assertions, assumptions, and loop invariants.
Transformation \tr[loop] relies on Soot's loop analysis capabilities
to identify loops in Vimp's unstructured control flow;
then, it expresses their invariants
by means of assertions and assumptions.
As shown in \autoref{fig:invariant-expansion},
\tr[loop] checks that the invariant holds upon loop entry
(label \Vi{head}),
at the end of each iteration (\Vi{head} again),
and at every exit point (label \Vi{exit}).

\begin{figure}
  \centering
  \begin{subfigure}[t]{0.36\linewidth}
\begin{lstlisting}[language=BBlib,numbers=none,frame=none]
k = 0;
while (k < 10) {
  invariant(lte(k, 10) & lte(k, X));
  k++;
  if (k >= X) break;
}
return k;
\end{lstlisting}
  \end{subfigure}
  \hfill
  \begin{subfigure}[t]{0.29\linewidth}
\begin{lstlisting}[language=Vimp,numbers=none,frame=none]
k := 0;
head:
  if k >= 10 goto exit;
  invariant k <= 10 && k <= X;
  k := k + 1;
  if k >= X goto exit;
back: goto head;
exit:
  return k;
\end{lstlisting}
  \end{subfigure}
  \hfill
  \begin{subfigure}[t]{0.31\linewidth}
\begin{lstlisting}[language=Vimp,numbers=none,frame=none]
k := 0;
head: assert k <= 10 && k <= X;
  if k >= 10 goto exit;
  assume k <= 10 && k <= X;
  k := k + 1;
  if k >= X goto exit;
back: goto head;
exit: assert k <= 10 && k <= X;
  return k;
\end{lstlisting}
  \end{subfigure}
  \caption{A loop in Java (left), its unstructured representation in Vimp (middle), and the transformation \tr[loop] of its invariant into assertions and assumptions (right).}
  \label{fig:invariant-expansion}
\end{figure}

\begin{figure}
  \centering
  \begin{subfigure}[t]{0.25\linewidth}
\begin{lstlisting}[language=BBlib,numbers=none,frame=none]
try {
  x = o.size();
  if (x == 0)
    throw new E();
} catch (E e) {
  x = 1;
}
\end{lstlisting}
  \end{subfigure}
  \hfill
  \begin{subfigure}[t]{0.25\linewidth}
\begin{lstlisting}[language=Jimple,numbers=none,frame=none]
$\ell_1$: x := o.size();


$\ell_2$: if x != 0 goto $\ell_5$;
$\ell_3$: e := new E();
$\ell_4$: throw e;


$\ell_5$: goto $\ell_6$;
hE:
  e := @caught;
  x := 1;
$\ell_6$: ...
\end{lstlisting}
  \end{subfigure}
  \hfill
  \begin{subfigure}[t]{0.43\linewidth}
\begin{lstlisting}[language=Vimp,numbers=none,frame=none]
$\ell_1$: x := o.size();
    if @thrown == void goto skip$_2$;
    $\tr[exc]($throw @thrown;$)$
skip$_2$: $\ell_2$: if x != 0 goto $\ell_5$;
$\ell_3$: e := new E();
$\ell_4$: @thrown := e;
    if $\;$!(@thrown instanceof E) goto skip$_5$;
    goto hE;
skip$_5$: $\ell_5$: goto $\ell_6$;
hE:
  e := @thrown; @thrown := void;
  x := 1;
$\ell_6$: ...
\end{lstlisting}
  \end{subfigure}
  \caption{A try-catch block in Java (left), its unstructured representation as a trap in Grimp (middle, empty lines are for readability), and its transformation \tr[exc] in Vimp with explicit exceptional control flow (right).}
  \label{fig:exception-cf}
\end{figure}

\subsection{Modeling Exceptional Control Flow}
\label{subsec:exceptional-control-flow}
\vspace{-2pt}
Bytecode stores a block's exceptional behavior
in a data structure called the \emph{exception table}.\urlcite{https://docs.oracle.com/javase/specs/jvms/se7/html/jvms-2.html\#jvms-2.10}
Soot represents each table entry as a \emph{trap},
which renders a try-catch block in Grimp bytecode.
Precisely, a trap $t$ is defined by:
\begin{enumerate*}
\item a block of instructions $B_t$
  that may throw exceptions;
\item the type $E_t$ of the handled exceptions;
\item a label $h_t$ to the handler instructions
  (which terminates with a jump back to the end of $B_t$).
\end{enumerate*}
When executing $B_t$ throws an exception whose type conforms to $E_t$,
control jumps to $h_t$.
At the beginning of the handler code,
Grimp introduces \Gr{e := @caught},
which stores into a local variable \Gr{e} of the handler
a reference \Gr{@caught} to the thrown exception object.
\autoref{fig:exception-cf} shows an example of try-catch block in Java
(left) and the corresponding trap in Grimp (middle):
$\ell_1, \ldots, \ell_5$ is the instruction block,
\Gr{E} is the exception type,
and \Gr{hE} is handler's entry label.
The rest of this section describes \byteback's transformation \tr[exc],
which transforms the implicit exceptional control flow of Grimp traps
into explicit control flow in Vimp.

\subsubsection{Explicit Exceptional Control-Flow}
Grimp's variable \Gr{@caught}
is called \Vi{@thrown} in Vimp.
While \Gr{@caught} is read-only in Grimp---where it only refers to the currently handled exception---\Vi{@thrown} can be assigned to in Vimp.
This is how \byteback makes exceptional control flow explicit:
assigning to \Vi{@thrown} an exception object \Vi{e}
signals that \Vi{e} has been thrown;
and setting \Vi{@thrown} to \Vi{void}
marks the current execution as normal.
Thus, \byteback's Vimp encoding sets \Vi{@thrown := void}
at the beginning of a program's execution,
and then manipulates the special variable
to reflect the bytecode semantics of exceptions
as we outline in the following.
With this approach, the Vimp encoding of a \J{try} block
simply results from encoding each of the block's instructions
explicitly according to their potentially exceptional behavior.

\subsubsection{Throw Instructions}
Transformation \tr[exc]
desugars \Gr{throw} instructions into
explicit assignments to \Vi{@thrown} and jumps
to the suitable handler.
A \Gr{throw e} instruction
within the blocks of $n$ traps $t_1, \ldots, t_n$---handling exceptions of types $E_1, \ldots, E_n$
with handlers at labels $h_1, \ldots, h_n$---is transformed into:
\begin{small}
\begin{align*}
  \tr[exc]\big(\Gr{throw e}\big) & =\
  \left(
  \begin{array}{ll}
    & \Vi{@thrown := e;}\\
    & \Vi{if !$\!\!$(@thrown instanceof $\ E_1$) goto skip$_1$;}\\
    & \Vi{goto $\ h_1$;}\\
    \Vi{skip$_1$:} &
                     \Vi{if !$\!\!$(@thrown instanceof $\ E_2$) goto skip$_2$;}\\
    & \Vi{goto $\ h_2$;}\\
    \Vi{skip$_2$:}
    & \Vi{if !$\!\!$(@thrown instanceof $\ E_3$) goto skip$_3$;} \\
    & \vdots \\
    \Vi{skip$_{n-1}$:}
    & \Vi{if !$\!\!$(@thrown instanceof $\ E_n$) goto skip$_n$;} \\
    & \Vi{goto $\ h_n$;}\\
    \Vi{skip$_n$:} & \Vi{return;  // propagate exception to caller }
  \end{array} \right)
\end{align*}
\end{small}
The assignment to \Vi{@thrown} stores a reference
to the thrown exception object \Vi{e};
then, a series of checks determine if \Vi{e}
has type that conforms to any of the handled exception types;
if it does, execution jumps to the corresponding handler.

Transformation \tr[exc] also replaces
the assignment \Gr{e := @caught} that Grimp puts at the beginning of every handler
with \Vi{e := @thrown; @thrown := void},
signaling that the current exception is handled,
and thus the program will resume normal execution.

\subsubsection{Exceptions in Method Calls}
A called method may throw an exception,
which the caller should propagate or handle.
Accordingly, transformation \tr[exc] adds after
every method call instructions to check whether the caller set
variable \Vi{@thrown} and, if it did, to handle the exception
within the caller as if it had been directly thrown by it.
\begin{small}
\begin{align*}
  \tr[exc]\big(\Gr{m($a_1, \ldots, a_m$)}\big) & =\
  \left(
  \begin{array}{ll}
    & \Vi{m($a_1, \ldots, a_m$);} \\
    & \Vi{if (@thrown == void) goto skip;} \\
    & \tr[exc]\left(\Vi{throw @thrown}\right) \\
    \Vi{skip:} & \Vi{/* code after call */}
  \end{array} \right)
\end{align*}
\end{small}

\subsubsection{Potentially Excepting Instructions}
\label{subsubsec:pei}
Some bytecode instructions may implicitly throw exceptions
when they cannot execute normally.
In \autoref{code:into}'s running example,
\J{r.read()} throws a \J{NullPointer} exception if \J{r} is \J{null};
and the assignment to \J{a[i]} throws an \J{IndexOutOfBounds} exception if \J{i}
is not between \J{0} and \J{a.length - 1}.
Transformation \tr[exc] recognizes such
potentially excepting instructions and
adds explicit checks that capture their implicit exceptional behavior.
Let \op be an instruction that throws an exception of type \op[E]
when condition \op[F] holds;
\tr[exc] transforms \op as follows.
\begin{small}
\begin{align*}
  \tr[exc]\big(\op\big) & =\
  \left(
  \begin{array}{ll}
    & \Vi{if !}\!\!\op[F]\ \;\Vi{goto normal;} \\
    & \tr[exc]\left(
      \begin{array}{l}
        \Vi{e := new}\ \op[E]\Vi{()}; \\
        \Gr{throw e;} \\
      \end{array}
      \right) \\
    \Vi{normal:} & \op
  \end{array} \right)
\end{align*}
\end{small}

By chaining multiple checks,
transformation \tr[exc] handles
instructions that may throw multiple implicit exceptions.
For example, here is how it encodes the potentially excepting semantics
of array lookup \J{a[i]},
which fails if \J{a} is \J{null} or \J{i} is out of bounds.
\begin{small}
\begin{align*}
  \tr[exc]\big(\Vi{res := a[i]}\big) & =\
  \left(
  \begin{array}{ll}
    & \Vi{if !}\!\!\Vi{(a == null) goto normal$_1$;} \\
    & \tr[exc]\left(
      \begin{array}{l}
      \Vi{e$_1$ := new NullPointerException();} \\
      \Gr{throw e$_1$;} \\
      \end{array}
      \right) \\
    \Vi{normal$_1$:}
    & \Vi{if !}\!\!\Vi{(0 <= i && i < a.length) goto normal$_2$;} \\
    & \tr[exc]\left(
      \begin{array}{l}
        \Vi{e$_2$ := new IndexOutOfBoundsException();} \\
        \Gr{throw e$_2$;} \\
      \end{array}
      \right) \\
    \Vi{normal$_2$:} & \Vi{res := a[i];}
  \end{array} \right)
\end{align*}
\end{small}

\subsection{Transformation Order and Boogie Code Generation}
\label{sec:trasformation-order}
\byteback applies the transformations \tr from Grimp to Vimp
in a precise order that incrementally encodes the full program semantics
respecting dependencies.

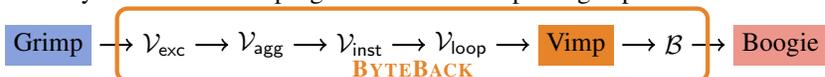
\begin{tikzpicture}
  \matrix[row sep=0mm,column sep=4.5mm] {
\node (grimp) {\colorbox{jimplecol}{\text{Grimp}}};
    &
    \node (exc) {\tr[exc]};
    &
    \node (agg) {\tr[agg]};
    &
    \node (inst) {\tr[inst]};
    &
    \node (loop) {\tr[loop]};
    &
    \node (vimp) {\colorbox{bbcol}{\text{Vimp}}};
    &
    \node (trb) {$\mathcal{B}$};
    &
    \node (boogie) {\colorbox{boogiecol}{\text{Boogie}}};
\\
  };

  \begin{scope}[->,thick]
    \draw (grimp) -- node (begin-bb) {} (exc);
    \draw (exc) -- (agg);
    \draw (agg) -- (inst);
    \draw (inst) -- (loop);
    \draw (loop) -- (vimp);
    \draw (vimp) -- (trb);
    \draw (trb) -- node (end-bb) {} (boogie);
  \end{scope}
  
  \node [fit=(begin-bb)(end-bb),draw=bbcol,
  ultra thick,rounded corners,label={[bbcol]south:\textbf{\byteback}},inner xsep=-4pt,inner ysep=10pt,outer ysep=-11.5pt] (byteback-detail) {};
  
\end{tikzpicture}

Like raw bytecode,
the source Grimp representation on which \byteback operates is a form of three-address code,
where each instruction performs exactly one operation
(a call, a dereferencing, or a unary or binary arithmetic operation).
\begin{enumerate*}
\item \byteback first applies \tr[exc]
  to make the exceptional control flow explicit.
\item Then, it aggregates expressions using \tr[agg].
\item Transformation \tr[inst] is applied next to every instruction;
in turn, \tr[inst] relies on transformations \tr[exp], \tr[stm], and \tr[type]
  (presented in \autoref{subsec:vimp}) to process the expressions
  and types manipulated by the instructions (the instructions themselves do not change, and hence \tr[inst]'s definition is straightforward).
\item Finally, it applies \tr[loop] to encode loop invariants as intermediate
  assertions; since this transformation is applied after \tr[exc],
  the loop invariants can be checked at all loop exit points---normal and exceptional.
\end{enumerate*}

The very last step $\mathcal{B}$ of \byteback's pipeline
takes the fully transformed Vimp program
and encodes it as a Boogie program.
Thanks to Vimp's design, and to the transformation \tr applied to Grimp,
the Vimp-to-Boogie translation is straightforward.
In addition, \byteback also generates
a detailed Boogie axiomatization of all logic functions
used to model various parts of JVM execution---which we described in greater detail in previous work~\cite{PF-FM23-ByteBack}.
One important addition is
an axiomatization of subtype relations among exception types,
used by Boogie's \B{instanceof} function
to mirror the semantics of the homonymous Java operator.
Consider the whole tree $T$ of exception types used in the program:\footnote{Since the root of all exception types in Java is \emph{class} \J{Throwable}---a concrete class---an exception type cannot be a subtype of multiple exception classes, and hence $T$ is strictly a tree.}
each node is a type, and its children are its direct subtypes.
For every node \B{C} in the tree,
\byteback produces one axiom asserting that every child \B{X} of \B{C}
is a subtype of \B{C} ($\MB{X} \subtype \MB{C}$),
and one axiom for every pair \B{X}, \B{Y} of \B{C}'s
children asserting that any descendant types $x$ of \B{X} and $y$ of \B{Y}
are \emph{not} related by subtyping (in other words, the subtrees rooted in \B{X} and \B{Y} are disjoint):
$
  \forall x, y\colon \MB{Type}\ \MB{::}\ 
  x \subtype \MB{X}
  \sand
  y \subtype \MB{Y}
  \Longrightarrow
  x \notsubtype y
  \sand
  y \notsubtype x
$.

\subsection{Implementation Details}
\label{subsec:impl}

\subsubsection{\bblib as Specification Language}
We are aware that \bblib's syntax and conventions may be inconvenient at times;
they were designed to deal with the fundamental constraints that any
specification must be expressible in the source code
and still be fully available for analysis in bytecode after compilation.
This rules out the more practical approaches (e.g., comments)
adopted by source-level verifiers.
More user-friendly notations could be introduced on top of \bblib ---but doing so is outside the present paper's scope.

\subsubsection{Attaching Annotations}
As customary in deductive verification,
\byteback models calls using the modular semantics,
whereby every called method needs a meaningful specification
of its effects within the caller.
To support more realistic programs
that call to Java's standard library methods,
\bblib supports the \BBl{@Attach} annotation:
a class \BBl{S} annotated with \BBl{@Attach(I.class)}
declares that any specification of any method in \J{S}
serves as specification
of any method with the same signature in \J{I}.
We used this mechanism to model the fundamental behavior of widely used methods
in Java's, Scala's, and Kotlin's standard libraries.
As a concrete example,
we specified that the constructors of common exception classes
do not themselves raise exceptions.

\subsubsection{Implicit Exceptions}
\autoref{subsec:exceptional-control-flow}
describes how \byteback
models potentially excepting instructions.
The mechanism is extensible,
and the current implementation supports
the ubiquitous \J{NullPointer} and \J{IndexOutOfBounds} exceptions, 
as shown in \autoref{tab:guarding-strategies}.
Users can selectively enable or disable
these checks for implicitly thrown exceptions
either for each individual method,
or globally for the whole program.

\begin{table}[!bt]
  \centering
  \setlength{\tabcolsep}{2pt}
  \begin{tabular}{l l l}
    \toprule
    \multicolumn{1}{c}{\textsc{excepting instructions}}
    & \multicolumn{1}{c}{\textsc{exception}}
    & \multicolumn{1}{c}{\textsc{condition}}         \\
    \midrule
    dereferencing $\J{o._}, \J{o[i]}$
    & \J{NullPointerException}
    & \Vi{o == null}  \\
    array access \J{a[i]}
    & \J{IndexOutOfBoundsException}
    & \Vi{!(0 <= i && i < a.length)} \\
    \bottomrule
  \end{tabular}
  \caption{Potentially excepting instructions currently supported by \byteback.}
  \label{tab:guarding-strategies}
\end{table}

\subsubsection{Dependencies}
\byteback is implemented as a command-line tool
that takes as input
a classpath and a set $E$ of class files within that path.
The analysis collects
all classes on which the entry classes in $E$ recursively depend---where ``$A$ depends on $B$'' means that
$A$ inherits from or is a client of $B$.
After collecting all dependencies,
\byteback feeds them through its verification toolchain (\autoref{fig:workflow})
that translates them to Boogie.
In practice, \byteback is configured with a list of \emph{system packages}---such as \J{java.lang}---that are treated differently:
their implementations are ignored
(i.e., not translated to Boogie for verification),
but their interfaces and any specifications
are retained to reason about their clients.
This makes the verification process more lightweight,

\subsubsection{Features and Limitations}
The main limitations of \byteback's previous version~\cite{PF-FM23-ByteBack}
were a lack of support for exception handling and \Gr{invokedynamic}.
As discussed in the rest of the paper,
\byteback now fully supports reasoning about exceptional behavior.
We also added a, still limited, support for \Gr{invokedynamic}:
any instance of \Gr{invokedynamic}
is conservatively treated as a call whose effects are unspecified;
furthermore,
we introduced ad hoc support to reason about concatenation and comparison
of string literal---which are implemented using \Gr{invokedynamic} since Java~9.\footnote{\scriptsize\url{https://docs.oracle.com/javase/9/docs/api/java/lang/invoke/StringConcatFactory.html}}
A full support of \Gr{invokedynamic} still belongs to future work.

Other remaining limitations of \byteback's current implementation
are a limited support of string objects,
and no modeling of numerical errors such as overflow
(i.e., numeric types are encoded with infinite precision).
Adding support for all of these features is
possible by extending \byteback's current approach.

\section{Experiments}
\label{sec:experiments}

We demonstrated \byteback's capabilities by running its
implementation on a collection of annotated programs
involving exceptional behavior in Java, Scala, and Kotlin.

\subsection{Programs}

\autoref{tab:experiments} lists the \n{count} programs
that we prepared for these experiments;
all of them involve \emph{some} exceptional behavior
in different contexts.\footnote{We focus on these exception-related programs,
  but the latest version of \byteback also verifies correctly
  the \n{count/non-exceptional} other programs we introduced in previous work
  to demonstrate its fundamental verification capabilities.
}
More than half of the programs ($\n{count/java}/\n{count}$)
are in Java:
\n{count/j8} only use language features that have been available since Java~8,
and another \n{count/j17} rely on more recent features available since Java~17.
To demonstrate how targeting bytecode
makes \byteback capable of verifying, at least in part, other JVM languages,
we also included \n{count/s2} programs written in Scala (version~2 of the language),
and \n{count/k18} programs written in Kotlin (version~1.8.0).
Each program/experiment consists of one or more classes with their dependencies,
which we annotated with \bblib to specify exceptional and normal behavior,
as well as other assertions needed for verification (such as loop invariants).
The examples total \n{total/SourceLinesOfCode} lines of code and annotations,
with hundreds of annotations and \n{total/MethodCount} methods
(including \bblib specification methods)
involved in the verification process.
According to their features,
the experiments can be classified into two groups:
feature experiments and algorithmic experiments.

\subsubsection{Feature Experiments}

Java~8 programs 
\autoref{ex:j8-first-feature}--\autoref{ex:j8-last-feature},
Java~17 program
\autoref{ex:j17-first-feature},
Scala programs
\autoref{ex:s2-first-feature}--\autoref{ex:s2-last-feature},
and Kotlin programs
\autoref{ex:k18-first-feature}--\autoref{ex:k18-last-feature}
are \emph{feature} experiments:
each of them exercises a small set
of exception-related language features;
correspondingly, their specifications
check that \byteback's verification process
correctly captures the source language's semantics of those features.
For example,
experiments
\autoref{ex:j8-throwcatch},
\autoref{ex:s2-throwcatch},
and \autoref{ex:k18-throwcatch}
feature different combinations
of try-catch blocks and throw statements
that can be written in Java, Scala, and Kotlin,
and test whether \byteback correctly reconstructs all possible
exceptional and normal execution paths that can arise.
A more specialized example is experiment~\autoref{ex:j8-throwcatch-loop},
which verifies the behavior of loops with both normal and exceptional
exit points.

\subsubsection{Algorithmic Experiments}
Java~8 programs
\autoref{ex:j8-first-algo}--\autoref{ex:j8-last-algo},
Java~17 programs 
\autoref{ex:j17-first-algo}--\autoref{ex:j17-last-algo},
Scala programs
\autoref{ex:s2-first-algo}--\autoref{ex:s2-last-algo},
and Kotlin programs 
\autoref{ex:k18-first-algo}--\autoref{ex:k18-last-algo}
are \emph{algorithmic} experiments:
they implement classic algorithm
that also use exceptions
to signal when their inputs are 
invalid.
The main difference between feature and algorithmic experiments
is specification:
algorithmic experiments usually have more complex
pre- and postconditions than feature experiments,
which they complement with specifications
of exceptional behavior on the corner cases.
For example, experiments
\autoref{ex:j8-array-reverse},
\autoref{ex:s2-array-reverse},
and \autoref{ex:k18-array-reverse}
implement array reversal algorithms;
their postconditions specify that the input array is correctly reversed;
and other parts of their specification say that
they result in an exception if the input array is null.
Experiment~\autoref{ex:j17-readresource}
is an extension of \autoref{fig:motivating-examples}'s
running example,
where the algorithm is a simple stream-to-array copy
implemented in a way that may give rise to various kinds of exceptional behavior.

Experiments \autoref{ex:j8-arraylist} and \autoref{ex:j8-linkedlist}
are the most complex programs in our experiments:
they include a subset of 
the complete implementations of Java's
\J{ArrayList} and \J{LinkedList} standard library classes,\urlcite{https://docs.oracle.com/javase/8/docs/api/java/util/package-summary.html}
part of which we annotated with basic postconditions
and a specification
of their exceptional behavior
(as described in their official documentation).
In particular,
\J{ArrayList}'s exceptional specification
focuses on possible failures of the class constructor
(for example, when given a negative number as initial capacity);
\J{LinkedList}'s specification focuses on
possible failures of some of the read methods
(for example, when trying to get elements from an empty list).
Thanks to \bblib's features
(including the \BBl{@Attach} mechanism described in \autoref{subsec:impl}),
we could add annotations without modifying the implementation of these classes.
Note, however, that we verified relatively simple specifications,
focusing on exceptional behavior;
a dedicated support for complex
data structure functional specifications~\cite{PTF-FAOC17,DBLP:journals/sttt/HiepMBBG22}
exceeds \bblib's current capabilities and belongs to future work.

\subsubsection{Implicit Exceptions}
As explained in \autoref{subsec:impl},
users of \byteback can enable or disable checking of implicitly thrown
exceptions.
Experiments
\autoref{ex:j8-npe}, \autoref{ex:j17-readresource},
\autoref{ex:s2-npe}, and \autoref{ex:k18-npe}
check implicit \J{null}-pointer exceptions;
experiments \autoref{ex:j8-iobe}, \autoref{ex:j17-readresource}, \autoref{ex:s2-iobe}, and \autoref{ex:k18-iobe}
check implicit out-of-bounds exceptions;
all other experiments do not use any implicitly thrown exceptions,
and hence we disabled the corresponding checks.

\begin{table}[!phtb]
  \setlength{\tabcolsep}{3pt}
  \renewcommand{\arraystretch}{0.45}
  \scriptsize
  \centering
  \begin{tabular}{rllrrrrrrrr}
    \toprule
    \multicolumn{1}{c}{\textsc{\#}}
 & \multicolumn{1}{c}{\textsc{experiment}}
 & \multicolumn{1}{c}{\textsc{lang}}
 & \multicolumn{1}{c}{\textsc{encoding}}
 & \multicolumn{1}{c}{\textsc{verification}}
 & \multicolumn{1}{c}{\textsc{source}}
 & \multicolumn{1}{c}{\textsc{boogie}}
 & \multicolumn{1}{c}{\textsc{met}}
 & \multicolumn{3}{c}{\textsc{annotations}}
    \\
    \cmidrule(lr){4-5}\cmidrule(lr){6-7}\cmidrule(lr){8-8}\cmidrule(lr){9-11}
 &
 & 
 & \multicolumn{2}{c}{\textsc{time} [s]}
 & \multicolumn{2}{c}{\textsc{size} [LOC]}

 & 
 & \multicolumn{1}{c}{$P$}
 & \multicolumn{1}{c}{$S$}
 & \multicolumn{1}{c}{$E$}
                                                                                            \\
    \midrule
    \exprowc\label{ex:j8-first-feature}\label{ex:j8-first}\label{ex:first}\label{ex:j8-iobe}
 & Implicit Index Out of Bounds & J~8   & \exprow{j8/exceptions/PotentialIndexOutOfBounds}  \\
    \exprowc\label{ex:j8-npe}
 & Implicit Null Dereference    & J~8   & \exprow{j8/exceptions/PotentialNullDereference}   \\
    \exprowc
 & Multi-Catch                  & J~8   & \exprow{j8/exceptions/MultiCatch}                 \\
    \exprowc\label{ex:j8-throwcatch}
 & Throw-Catch                  & J~8   & \exprow{j8/exceptions/Basic}                      \\
    \exprowc\label{ex:j8-throwcatch-loop}
 & Throw-Catch in Loop          & J~8   & \exprow{j8/exceptions/Loop}                       \\
    \exprowc
 & Try-Finally                  & J~8   & \exprow{j8/exceptions/TryFinally}                 \\
    \exprowc\label{ex:j8-last-feature}
 & Try-With-Resources           & J~8   & \exprow{j8/exceptions/TryWithResources}           \\
    \exprowc\label{ex:j8-array-reverse}\label{ex:j8-first-algo}
 & Array Reverse                & J~8   & \exprow{j8/algorithm/ArrayReverse}                \\
    \exprowc
 & Binary Search                & J~8   & \exprow{j8/algorithm/BinarySearch}                \\
    \exprowc\label{ex:j8-gcd}
 & GCD                          & J~8   & \exprow{j8/algorithm/GCD}                         \\
    \exprowc
 & Linear Search                & J~8   & \exprow{j8/algorithm/LinearSearch}                \\
    \exprowc
 & Selection Sort (\J{double})  & J~8   & \exprow{j8/algorithm/DoubleSelectionSort}         \\
    \exprowc
 & Selection Sort (\J{int})     & J~8   & \exprow{j8/algorithm/IntegerSelectionSort}        \\
    \exprowc
 & Square of Sorted Array       & J~8   & \exprow{j8/algorithm/SquareSortedArray}           \\
    \exprowc
 & Sum                          & J~8   & \exprow{j8/algorithm/IntegerSum}                  \\
    \exprowc\label{ex:j8-first-library}\label{ex:j8-arraylist}
 & \J{ArrayList}                & J~8   & \exprow{j8/library/java/util/ArrayList}           \\
\exprowc\label{ex:j8-last-library}\label{ex:j8-last-algo}\label{ex:j8-last}\label{ex:j8-linkedlist}
 & \J{LinkedList}               & J~8   & \exprow{j8/library/java/util/LinkedList}          \\
\exprowc\label{ex:j17-first-feature}\label{ex:j17-last-feature}\label{ex:j17-first}
 & Try-With-Resources on Local  & J~17  & \exprow{j17/exceptions/TryWithResources}          \\
    \exprowc\label{ex:j17-first-algo}\label{ex:j17-summary}
 & Summary                      & J~17  & \exprow{j17/examples/Summary}                    \\
    \exprowc\label{ex:j17-last-algo}\label{ex:j17-readresource}
 & Read Resource                & J~17  & \exprow{j17/examples/ReadResource}                \\
    \exprowc\label{ex:s2-first-feature}\label{ex:s2-iobe}
 & Implicit Index Out of Bounds & S~2   & \exprow{s2/exceptions/PotentialIndexOutOfBounds}  \\
    \exprowc\label{ex:s2-npe}
 & Implicit Null Dereference    & S~2   & \exprow{s2/exceptions/PotentialNullDereference}   \\
    \exprowc
 & Multi-Catch                  & S~2   & \exprow{s2/exceptions/MultiCatch}                 \\
    \exprowc\label{ex:s2-throwcatch}
 & Throw-Catch                  & S~2   & \exprow{s2/exceptions/Basic}                      \\
    \exprowc\label{ex:s2-last-feature}
 & Try-Finally                  & S~2   & \exprow{s2/exceptions/TryFinally}                 \\
    \exprowc\label{ex:s2-array-reverse}\label{ex:s2-first-algo}
 & Array Reverse                & S~2   & \exprow{s2/algorithm/ArrayReverse}                \\
    \exprowc
 & Counter                      & S~2   & \exprow{s2/instance/Counter}                      \\
    \exprowc\label{ex:s2-gcd}
 & GCD                          & S~2   & \exprow{s2/algorithm/GCD}                         \\
    \exprowc\label{ex:s2-last-algo}\label{ex:s2-linearsearch}
 & Linear Search                & S~2   & \exprow{s2/algorithm/LinearSearch}                \\
    \exprowc\label{ex:k18-first-feature}\label{ex:k18-iobe}
 & Implicit Index Out of Bounds & K~1.8 & \exprow{k18/exceptions/PotentialIndexOutOfBounds} \\
    \exprowc\label{ex:k18-npe}
 & Implicit Null Dereference    & K~1.8 & \exprow{k18/exceptions/PotentialNullDereference}  \\
    \exprowc\label{ex:k18-throwcatch}
 & Throw-Catch                  & K~1.8 & \exprow{k18/exceptions/Basic}                     \\
    \exprowc\label{ex:k18-last-feature}
 & Try-Finally                  & K~1.8 & \exprow{k18/exceptions/TryFinally}                \\
    \exprowc\label{ex:k18-array-reverse}\label{ex:k18-first-algo}
 & Array Reverse                & K~1.8 & \exprow{k18/algorithm/ArrayReverse}              \\
    \exprowc
 & Counter                      & K~1.8 & \exprow{k18/algorithm/Counter}                   \\
    \exprowc\label{ex:k18-gcd}
 & GCD                          & K~1.8 & \exprow{k18/algorithm/GCD}                       \\
    \exprowc\label{ex:k18-last-algo}\label{ex:k18-linearsearch}
 & Linear Search                & K~1.8 & \exprow{k18/algorithm/LinearSearch}              \\
    \midrule
 & \textbf{total}
 & 
 & \n[1]{total/ConversionTime}[0.001]
 & \n[1]{total/VerificationTime}[0.001]
 & \n{total/SourceLinesOfCode}
 & \n{total/BoogieLinesOfCode}
 & \n{total/MethodCount}
 & \n{total/SpecPredicateCount}
 & \n{total/SpecFunctionalCount}
 & \n{total/SpecExceptionCount}                                                              \\[2pt]
 & \textbf{average}
 & 
 & \n[1]{average/ConversionTime}[0.001]
 & \n[1]{average/VerificationTime}[0.001]
 & \n[0]{average/SourceLinesOfCode}
 & \n[0]{average/BoogieLinesOfCode}
 & \n[0]{average/MethodCount}
 & \n[0]{average/SpecPredicateCount}
 & \n[0]{average/SpecFunctionalCount}
 & \n[0]{average/SpecExceptionCount}                                                         \\
    \bottomrule
  \end{tabular}
  \caption{Verification experiments with exceptional behavior used to demonstrate \byteback's capabilities.
    Each row reports: a numeric identifier \textsc{\#} and
    a short description of the \textsc{experiment};
    the source programming \textsc{lang}uage
    (\underline{J}ava~8, \underline{J}ava~17,
    \underline{S}cala~2, \underline{K}otlin~1.8);
    the wall-clock time (in seconds) taken for \textsc{encoding}
    bytecode into Boogie,
    and for the \textsc{verification} of the Boogie program;
    the size (in non-empty lines of code)
    of the \textsc{source} program with its annotations,
    and of the generated \textsc{boogie} program;
    the number of \textsc{met}hods
    that make up the program and its \bblib specification;
    and the number of \textsc{annotations} introduced for verification, 
    among:
    specification predicates $P$ (\BBl{@Predicate}),
    pre- and postconditions $S$ (\BBl{@Require}, \BBl{@Ensure}),
    and exception annotations $E$ (\BBl{@Raise}, \BBl{@Return}).
}
  \label{tab:experiments}
\end{table}

\subsection{Results}

All experiments ran on a Fedora~36 GNU/Linux machine with an Intel Core~i9-12950HX CPU (4.9GHz), running Boogie 2.15.8.0, Z3 4.11.2.0, and Soot 4.3.0.
To account for measurement noise,
we repeated the execution of each experiment \spellout{5} times
and report the average wall-clock running time of each experiment,
split into \byteback bytecode-to-Boogie encoding
and Boogie verification of the generated Boogie program.
We ran Boogie with default options
except for experiment~\autoref{ex:j17-summary},
which uses the \texttt{/infer:j} option
(needed to derive the loop invariant of the enhanced \J{for} loop,
whose index variable is implicit in the source code).

All of the experiments verified successfully.
To sanity-check that the axiomatization
or any other parts of the encoding introduced by \byteback
are consistent,
we also ran Boogie's so-called smoke test
on the experiments;\footnote{Smoke tests provide no absolute guarantee of consistency, but are often practically effective.}
these tests inject \B{assert false} in reachable parts of a Boogie program,
and check that none of them pass verification.

As you can see in \autoref{tab:experiments},
\byteback's running time is usually below 1.5 seconds;
and so is Boogie's verification time.
Unsurprisingly, programs~\autoref{ex:j8-arraylist} and~\autoref{ex:j8-linkedlist}
are outliers,
since they are made of larger classes with many dependencies;
these slow down both \byteback's encoding process
and
Boogie's verification,
which have to deal with many annotations and procedures to analyze and verify.

Only about \n[0]{average/ConversionOverhead}[100]|
of the time listed under \autoref{tab:experiments}'s column \textsc{encoding}
is taken by \byteback's actual encoding;
the rest is spent to perform class resolution
(\autoref{subsec:impl}) and to initialize Soot's analysis---which dominate \byteback's overall running time.

\section{Related Work}
\label{sec:related-work}

The state-of-the-art deductive verifiers for Java
include
OpenJML~\cite{OpenJML},
KeY~\cite{KeY},
and Krakatoa~\cite{Krakatoa};
they all process the source language directly,
and use variants of JML specification language---which offers support for specifying exceptional behavior.

\subsubsection{Exceptional Behavior Specifications}
Unlike \bblib,
where postconditions can refer to both exceptional and normal behavior,
JML clearly separates between the two, using 
\JML{ensures} and \JML{signals} clauses
(as demonstrated in \autoref{fig:jml-bblib-spec-comparison}).
These JML features are supported by
OpenJML, KeY, and Krakatoa according to their intended semantics. 
\begin{figure}[!h]
  \begin{subfigure}[t]{0.55\linewidth}
\begin{lstlisting}[language={[JML]Java},numbers=none]
//@ ensures this.a == a;
//@ signals (Throwable) this.a == a;
public void m(int a)
{ this.a = a; if ($\epsilon$) throw new RuntimeException(); }
\end{lstlisting}
    \label{code:jml-spec-comparison}
  \end{subfigure}
\hfill
  \begin{subfigure}[t]{0.4\linewidth}
\begin{lstlisting}[language=BBlib,numbers=none]
@Ensure(this.a == a)
public void m(int a)
{ this.a = a;
  if ($\epsilon$) throw new RuntimeException(); }
\end{lstlisting}
    \label{code:bblib-spec-comparison}
  \end{subfigure}
  \caption{Equivalent exceptional specifications in JML (left) and \bblib (right).}
  \label{fig:jml-bblib-spec-comparison}
\end{figure}

\subsubsection{Implicit Exceptional Behavior}
Implicitly thrown exceptions,
such as those occurring when accessing an array with an out-of-bounds index,
may be handled in different ways by a verifier:
\begin{enumerate*}
\item \label{es:ignore} ignore such exceptions;
\item \label{es:prohibit} implicitly check that such exceptions never occur;
\item \label{es:specify} allow users to specify these exceptions like explicit ones.
\end{enumerate*}
OpenJML and Krakatoa~\cite{KrakatoaJavaCard} follow strategy \ref{es:prohibit},
which is sound but loses some precision since it won't verify some
programs (such as \autoref{sec:examples}'s example);
KeY offers options to select any of these strategies,
which gives the most flexibility;
\byteback offers options \ref{es:ignore}
and \ref{es:specify},
so that users can decide
how thorough the analysis of exceptional behavior should be.

\subsubsection{Java Exception Features}
OpenJML, KeY, and Krakatoa~\cite{Krakatoa}
all support try-catch-finally blocks,
which have been part of Java since its very first version.
The first significant extension to exceptional feature occurred
with Java~7,
which introduced
multi-catch and try-with-resources blocks.\urlcite{https://www.oracle.com/java/technologies/javase/jdk7-relnotes.html}
KeY and Krakatoa support earlier versions of Java,
and hence they cannot handle either feature.
OpenJML supports many features of Java up to version~8,
and hence can verify programs using multi-catch or try-with-resources---with the exception of try-with-resources using an existing \J{final} variable,
a feature introduced only in Java~9.
As usual, our point here is not to criticize these state-of-the-art verification tools, but to point out how handling
the proliferation of Java language features
becomes considerably easier when targeting bytecode following \byteback's approach.

\subsubsection{Intermediate Representation Verifiers}
A different class of verifiers---including 
JayHorn~\cite{JayHorn,JayHorn2}, 
SeaHorn~\cite{SeaHorn},
and SMACK~\cite{SMACK}---target intermediate representations
(JVM bytecode for JayHorn, and LLVM bitcode for SeaHorn and SMACK).
Besides this similarity,
these tools' capabilities
are quite different from \byteback's:
they implement analyses based on model-checking
(with
verification conditions expressible as constrained Horn clauses,
or other specialized logics),
which provide a high degree of automation
(e.g., they do not require loop invariants)
to verify simpler, lower-level properties
(e.g., reachability).
Implicitly thrown exceptions are within the purview of
tools like JayHorn, which injects checks before each instruction
that may dereference a null pointer, access an index out of bounds,
or perform an invalid cast.
In terms of usage,
this is more similar to a specialized static analysis tool
that checks the absence of certain runtime errors~\cite{Infer,NullAway,CheckerFramework}
than to fully flexible,
but onerous to use,
deductive verifiers like \byteback.

BML~\cite{BML-tools} is a specification language for bytecode;
since it is based on JML, it is primarily used as a way of
expressing a high-level Java behavioral specification at the bytecode level.
This is useful for approaches
to proof-carrying code~\cite{pcc}
and proof transformations~\cite{proof-transf},
where one verifies a program's source-code
and then certifies its bytecode compilation
by directly transforming the proof steps.

\section{Conclusions}
\label{sec:conclusions}

Reasoning about exceptional behavior
at the level of Java bytecode
facilitates handling exception-related features
in any version of Java, as well as
in other JVM languages like Scala and Kotlin.
More generally,
the \byteback approach that we extended in this paper
can complement the core work in source-level deductive verification
and make it readily available to the latest languages and 
features.

\bibliographystyle{splncs04}
\bibliography{bb-exceptions}
\printendnotes[custom]

\end{document}
